\documentclass{emulateapj}

\newcommand{\citepeg}[1]{\citep[{e.g.,}][]{#1}}
\def\Swift{{\textit{Swift}}\,}

\begin{document}

\title{GRB 080503: Implications of a Naked Short Gamma-Ray Burst 
Dominated by Extended Emission}

\def\berk{1}
\def\hertfordshire{2}
\def\gsfc{3}
\def\umbc{4}
\def\ucsc{5}
\def\warwick{6}
\def\sloan{7}
\def\exeter{8}
\def\aao{9}
\def\chicago{10}
\def\swinburne{11}
\def\york{12}
\def\ssl{13}
\def\kavli{14}
\def\chile{15}
\def\denver{16}
\def\leicester{17}

\author{D.~A.~Perley\altaffilmark{\berk},
        B.~D.~Metzger\altaffilmark{\berk},
        J.~Granot\altaffilmark{\hertfordshire},
        N.~R.~Butler\altaffilmark{\berk},
        T.~Sakamoto\altaffilmark{\gsfc,\umbc},
        E.~Ramirez-Ruiz\altaffilmark{\ucsc},
        A.~J.~Levan\altaffilmark{\warwick},
        J.~S.~Bloom\altaffilmark{\berk,\sloan},
        A.~A.~Miller\altaffilmark{\berk},
        A.~Bunker\altaffilmark{\exeter,\aao},
        H.-W.~Chen\altaffilmark{\chicago},
        A.~V.~Filippenko\altaffilmark{\berk},
        N.~Gehrels\altaffilmark{\gsfc},
        K.~Glazebrook\altaffilmark{\swinburne},
        P.~B.~Hall\altaffilmark{\york},
        K.~C. Hurley\altaffilmark{\ssl},
        D.~Kocevski\altaffilmark{\kavli},
        W.~Li\altaffilmark{\berk},
        S.~Lopez\altaffilmark{\chile},
        J.~Norris\altaffilmark{\denver},
        A.~L.~Piro\altaffilmark{\berk},
        D.~Poznanski\altaffilmark{\berk},
        J.~X.~Prochaska\altaffilmark{\ucsc},
        E.~Quataert\altaffilmark{\berk},
        N.~Tanvir\altaffilmark{\leicester}
        }

 \altaffiltext{\berk}{Department of Astronomy, University of California, Berkeley, CA 94720-3411.}
 \altaffiltext{\hertfordshire}{Centre for Astrophysics Research, University of Hertfordshire, College Lane, Hatfield, Herts, AL10 9AB, UK.}
 \altaffiltext{\gsfc}{NASA Goddard Space Flight Center, Greenbelt, MD 20771.}
 \altaffiltext{\umbc}{CRESST/Joint Center for Astrophysics, University of Maryland, Baltimore County, Baltimore, MD 21250.}
 \altaffiltext{\ucsc}{Department of Astronomy and Astrophysics, UCO/Lick Observatory, University of California, 1156 High Street, Santa Cruz, CA 95064.}
 \altaffiltext{\warwick}{Department of Physics, University of Warwick, Coventry CV4 7AL, UK.}
 \altaffiltext{\sloan}{Sloan Research Fellow.}
 \altaffiltext{\exeter}{School of Physics, University of Exeter, UK}    
 \altaffiltext{\aao}{Anglo-Australian Observatory}
 \altaffiltext{\chicago}{Department of Astronomy and Astrophysics, University of Chicago, 5640 S. Ellis Ave, Chicago, IL 60637.}
 \altaffiltext{\swinburne}{Swinburne University of Technology}
 \altaffiltext{\york}{Department of Physics and Astronomy, York University, 4700 Keele St., Toronto, Ontario M3J 1P3, Canada}
 \altaffiltext{\ssl}{University of California, Berkeley, Space Sciences Laboratory, 7 Gauss Way, Berkeley, CA 94720-7450}
 \altaffiltext{\kavli}{Kavli Institute for Particle Astrophysics and Cosmology, Stanford University, 2575 Sand Hill Road M/S 29, Menlo Park, CA 94025}
 \altaffiltext{\chile}{Departamento de Astronom\'ia, Universidad de Chile, Casilla 36-D, Santiago, Chile}
 \altaffiltext{\denver}{University of Denver, Physics and Astronomy Department, Denver, CO 80208 }
 \altaffiltext{\leicester}{Department of Physics and Astronomy, University of Leicester, University Road, Leicester LE1 7RH}
 
\slugcomment{Submitted to ApJ 2008-11-07, accepted 2009-02-11.}

\begin{abstract}

We report on observations of GRB\ 080503, a short gamma-ray burst with
very bright extended emission (about 30 times the gamma-ray fluence of
the initial spike) in conjunction with a thorough comparison to other
short \Swift events.  In spite of the prompt-emission brightness,
however, the optical counterpart is extraordinarily faint, never
exceeding 25 mag in deep observations starting at $\sim$1 hr after the
BAT trigger.  The optical brightness peaks at $\sim 1\;$day and then
falls sharply in a manner similar to the predictions of
\citet{LiPaczynski1998} for supernova-like emission following
compact-binary mergers.  However, a shallow spectral index
and similar evolution in X-rays inferred from \textit{Chandra}
observations are more consistent with an afterglow interpretation.  
The extreme faintness of this probable afterglow relative to the 
bright gamma-ray emission argues for a very low-density medium 
surrounding the burst (a ``naked'' GRB), consistent with the lack of
a coincident host galaxy down to 28.5 mag in deep {\it Hubble Space
Telescope} imaging.  The late optical and X-ray peak
could be explained by a slightly off-axis jet or by a refreshed shock.
Our observations reinforce the notion that short gamma-ray bursts
generally occur outside regions of active star formation, but
demonstrate that in some cases the luminosity of the extended prompt
emission can greatly exceed that of the short spike, which may
constrain theoretical interpretation of this class of events.  This
extended emission is not the onset of an afterglow, and its relative
brightness is probably either a viewing-angle effect or intrinsic to
the central engine itself.  Because most previous BAT short bursts
without observed extended emission are too faint for this signature 
to have been detectable even if it were present at typical level, 
conclusions based solely on the observed presence or absence of
extended emission in the existing \Swift sample are premature.
\end{abstract}


\section{Introduction}
\label{sec:intro}

Despite significant progress since the launch of the \Swift satellite
\citep{Gehrels+2004}, the origin of short-duration, hard-spectrum gamma-ray
bursts (SGRBs) remains elusive.  Evidence has been available since the
early 1990s that SGRBs constitute a separate class from longer
GRBs on the basis of a bimodal distribution in duration \citep{Mazets+1981,Norris+1984} 
and spectral hardness \citep{Kouveliotou+1993}.  The supposition
that this phenomenological divide is symptomatic of a true physical
difference in the origin of the events was supported by the first
successful localizations of SGRB afterglows with the \Swift
X-ray telescope \citep{Burrows+2005} coincident with or apparently
very near low-redshift ($z < 0.5$) galaxies \citep{Gehrels+2005,Fox+2005}.  
Several of these galaxies clearly lack significant recent star formation
\citepeg{Prochaska+2006,Gorosabel+2006,Berger+2005}, many events
appeared at large offset from the candidate host
\citep{Bloom+2006,Bloom+2007,Stratta+2007}, and in some cases the
appearance of a bright supernova was definitively ruled out
\citepeg{Hjorth+2005a}.  All of these circumstantial clues seem to
suggest \citep{LeeRR2007,Nakar2007} a progenitor very different from the one
responsible for long-duration GRBs (LGRBs), which are predominately due to the
deaths of massive stars (see \citealt{WoosleyBloom2006} for a review).

The generally favored interpretation of SGRBs is the merger of two
highly compact degenerate objects: two neutron stars 
(NS--NS, \citealt{Eichler+1989,Meszaros+1992,Narayan+1992}) or a neutron
star and a black hole (NS--BH, \citealt{Paczynski+1991,Narayan+1992,Mochkovitch+1993,Kluzniak+1998,Janka+1999}).
However, other progenitor models \citepeg{MacFadyen+2005,Metzger+2008}
can also be associated with galaxies having low star-formation rates
(SFRs), and many SGRBs have also been associated with relatively
low-luminosity, high-SFR galaxies
\citep{Fox+2005,Hjorth+2005b,Covino+2006,Levan+2006} and at much higher redshifts
\citep{Berger+2007,Cenko+2008} than the better-known elliptical hosts
of the first few well-localized SGRBs 050509B and 050724.
(A review of SGRB progenitor models is given by \citealt{LeeRR2007}.)

In addition, even the conventional distinction between SGRBs and LGRBs
has been called into question by some recent events which poorly
conform to the traditional classification scheme.  A large number of
\Swift events which initially appeared to be ``short'' (based only
on the analysis of the first, most intense pulse) were then followed by 
an additional episode of long-lasting emission with a duration of up to
100~s or longer.  GRB~050724, which unambiguously occurred in an
elliptical host, is a member of this class, creating a breakdown in
the use of duration (in particular $T_{90}$, \citealt{Kouveliotou+1993})
as a classification criterion.  To further
complicate the picture, long GRB~060614 exploded in a very low-SFR
dwarf galaxy at $z=0.125$ and despite an intensive follow-up campaign
showed no evidence for a supernova, even if extremely underluminous 
($M_V > -12.3$, \citealt{GalYam+2006}).
Similar confusion clouds the physical origin of GRB~060505, which is
of long duration ($T_{90} = 4 \pm 1$ s) and occurred in a star-forming
region of a spiral galaxy \citep{Thoene+2008}, but also lacked supernova emission to very
deep limits \citep{Fynbo+2006}.  Two earlier bursts, XRF~040701
\citep{Soderberg+2005} and GRB~051109B \citep{GCN5387}, may constitute
additional examples of this subclass, though available limits in each
case are much shallower and the alternate possibility of host-galaxy
extinction is poorly constrained compared to the 2006 events.  On the
basis of these results and others, \citet{Zhang+2007} have called for
a new terminology for classification that does not refer to ``short''
and ``long'' but rather to Type I and Type II GRBs, in recognition of
the fact that duration alone is likely to be an imperfect proxy for
physical origin (see also \citealt{Gehrels+2006}, \citealt{Bloom+2008}, \citealt{Kann+2008}).

The true ``smoking gun'' for the merger model, the detection of
gravitational waves, is unlikely to occur before the completion of the
next generation of gravity-wave detectors, as the sensitivity of
current detectors (LIGO, \citealt{Abbott+2004}; 
and Virgo, \citealt{Acernese+2004}) is several orders of magnitude
below what would be necessary to detect a merger at what appears to be
a ``typical'' short GRB redshift of 0.2--1.0 \citep{Abbott+2008}.
However, degenerate-merger models do offer additional observationally
verifiable predictions.  

First, merger progenitors are much older than
massive stars and can travel far from their birthsites, especially if
they are subject to kicks which in some cases could eject the binary
system progenitor from the host galaxy entirely
\citep{Fryer+1999,Bloom+1999}.  Observationally, this should manifest 
itself in the form of large angular offsets between the burst position
and the host galaxy or even the lack of any observable host at all.
Such an trend has indeed been noted for many events \citepeg{Prochaska+2006}.  
The second prediction, however, has yet to be demonstrated:
if some SGRBs explode in galactic halos, then the extremely low associated 
interstellar density will result in a much fainter afterglow associated 
with the external shock:  a ``naked'' gamma-ray burst.  And while the 
afterglows of SGRBs tend to be fainter in an absolute sense
\citep{Kann+2008}, relative to the gamma-ray emission (on average, SGRBs have much lower
total fluences than long LGRBs) there appears to be no obvious difference between
SGRB and LGRB afterglows \citep{Nysewander+2008}.  Part of this may be a selection effect,
but the brightest SGRBs to date have all been associated with bright afterglows 
and cannot be ``naked''.

Second, during the merger process, a
significant amount of neutron-rich ejecta 
(including $\sim 10^{-3} {\rm M}_{\odot}$ of radioactive Ni, \citealt{Metzger+2008}) 
is believed to be ejected at
nonrelativistic velocities into interstellar space.  Nucleosynthesis
in this matter and the resulting radioactive decay would be expected
to produce a relatively long-lived optical counterpart, similar to
ordinary supernovae \citep{LiPaczynski1998}.
Unfortunately, the luminosity of the transient is generally much lower
and the timescale of evolution is significantly faster than in a
classical supernova.  Detection of this signature remains one of the
holy grails in the study of GRBs, though deep early limits for some
SGRBs have allowed some limits to be set on the physical
parameters of this phenomenon \citep{Bloom+2006,Hjorth+2005a,Kann+2008}.

In this paper, we present results from our follow-up campaign of
GRB\,080503, which we argue in \S\ref{sec:bat} is a prominent example
of the emerging subclass of SGRBs with extended episodes of bright, 
long-lasting prompt emission following the initial short spike.  In
\S\ref{sec:uvot}--\S\ref{sec:chandra} we present additional
space-based and ground-based observations of the event highlighting
several extreme and unusual features of this burst, including 
extreme optical faintness, a late light-curve peak, and a
very deep late-time limit on any coincident host galaxy.   
In \S\ref{sec:model} we attempt to interpret the observed 
behavior in the context of existing models of emission from 
GRB internal shocks, an unusual afterglow, and from mini-SN 
light, arguing that the latter is probably not a large
contributor at any epoch.  Finally, in \S\ref{sec:conclusions} 
we discuss the implications  of this event for GRB classification, 
and on the difficulties faced by future searches for mini-SN light 
associated with SGRBs.

\section{Observations}
\label{sec:obs}

\subsection{BAT Analysis and High-Energy Classification}
\label{sec:bat}

\begin{figure}
\centerline{
\includegraphics[scale=0.5,angle=0]{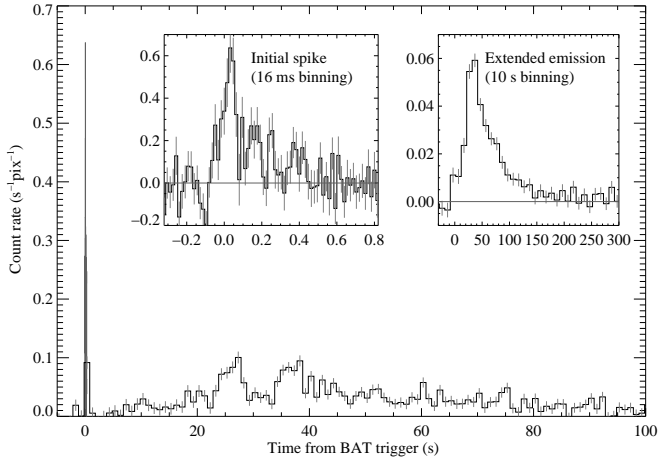}}
\caption{The BAT light curve of GRB 080503 with 1 s binning in the
15--150 keV band, with a 16~ms binning curve superposed for the duration
of the short spike near $t = 0$.  The short spike is also shown alone 
in the left inset.  An extended, highly-binned (10~s) light curve is shown
in the right inset, demonstrating the faint emission continuing until
about 200~s.}
\label{fig:bat_lc}
\end{figure}

\begin{figure}
\centerline{
\includegraphics[scale=0.6,angle=0]{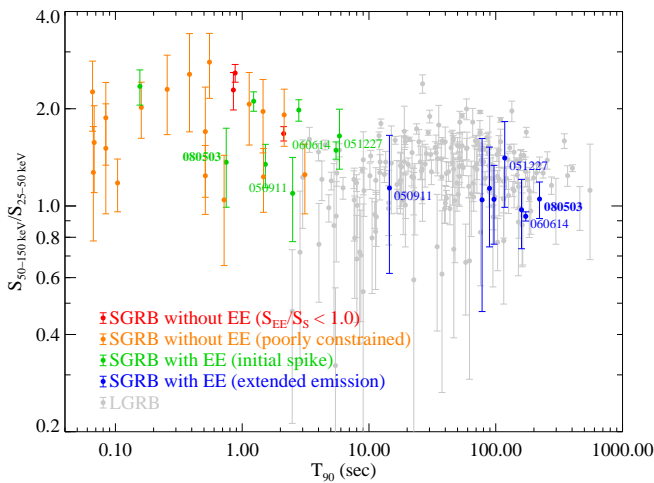}}
\caption{Duration-hardness plot for bursts detected by the \Swift BAT.
Long bursts are shown in gray.  Short bursts ($T_{90} < 2$ sec) are
colored based on the presence or absence of extended emission:
bursts without extended emission are shown in red, faint bursts
for which the presence of extended emission is poorly constrained
are orange, and short bursts with observed extended-emission
(including GRBs 050911, 060614, and 051227, whose classifications are
controversial) are plotted with the short spike (green) shown
separately from the extended emission (blue).  The $T_{90}$s and
hardness ratios measured for short-hard spikes in this population,
including GRB\ 080503, are generally consistent with those measured
for short bursts without extended emission. GRBs 060614 and 051227 may be
consistent with both classes, but are unusually long compared to any
short burst without extended emission. The extended-emission
components of all three events display similar hardness and duration
as the extended components of more traditional extended-emission
events, which form a tight cluster (GRB\ 050911 is an outlier).  In
general, however, the hardness in the \Swift channels is not a strong
criterion for classification \citep{Sakamoto+2006,Ohno+2008}.}
\label{fig:harddur}
\end{figure}

The \Swift Burst Alert Telescope (BAT) detected GRB\,080503 at
12:26:13 on 2008 May 3 (UT dates and times are used throughout this
paper).  The GRB light curve (Figure \ref{fig:bat_lc}) is a classic
example of a short GRB with extended emission: a short, intense initial 
spike with a duration of less than 1~s followed by a long episode of
extended emission starting at $\sim$10~s and lasting for
several minutes.  The overall $T_{90}$ for the entire event is 232~s.

\begin{figure}
\centerline{
\includegraphics[scale=0.5,angle=0]{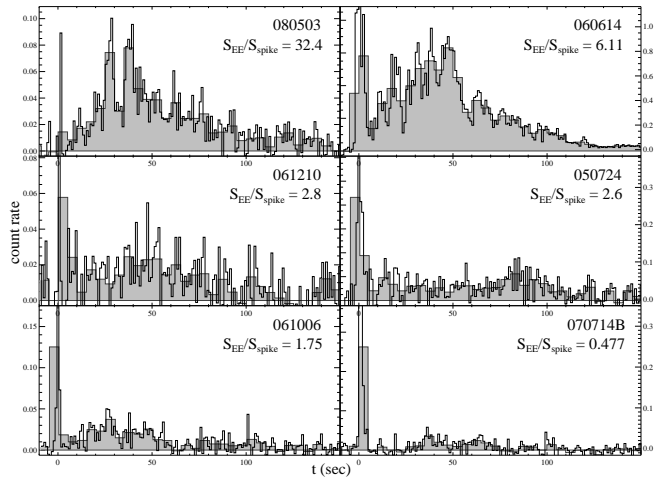}}
\caption{BAT 25--100 keV light curves of several different \Swift
short bursts with high signal-to-noise (S/N) extended emission,
including GRB\ 080503 (top left), showing the similar morphology of
these events.  The 1~s binned curve is plotted as a black line; a 5~s
binning is plotted in solid gray to more clearly show the
longer-duration extended emission which for most events is near the
detection threshold.  Possible short GRB\ 060614 is also shown; it
appears very similar to GRB\ 080503 except that the initial pulse is
significantly longer.}
\label{fig:multilc}
\end{figure}

Similar extended emission has been seen before in many short bursts
detected by both \Swift and BATSE (Figure \ref{fig:multilc}).  All
such events to date have remarkably similar general morphologies.
However, the fact that the long component is so dominant in this case
(factor of $\sim$30 in total fluence) raises the question of whether
this is truly a ``short'' (or Type I) GRB and not an event more
akin to the traditional LGRBs (Type II) in disguise.  To this end we
have reanalyzed the BAT data in detail and applied additional
diagnostics to further investigate the nature of this event.  We also
downloaded and re-analyzed BAT data from all other SGRBs (and
candidate SGRBs) with and without extended emission through the end of
2007.  A summary of the results of our analysis is presented in Table
\ref{tab:battable}.

The BAT data analysis was performed using the \textit{Swift} HEAsoft
6.5 software package.  The burst pipeline script, {\tt batgrbproduct},
was used to process the BAT event data.  In addition to the script, we
made separate spectra for the initial peak and the extended emission
interval by {\tt batbinevt}, applying {\tt batphasyserr} to the PHA
files.  Since the spectral interval of the extended emission includes
the spacecraft slew period, we created the energy response files for
every 5~s period during the time interval, and then weighted these
energy response files by the 5~s count rates to create the averaged
energy response.  The averaged energy response file was used for the
spectral analysis of the extended emission interval.  Similar methods
were employed for previous \Swift SGRBs.

For GRB\ 080503, the $T_{90}$ durations of the initial short spike and
the total emission in the 15--150 keV band are 0.32 $\pm$ 0.07 s, and
232 s respectively.  The peak flux of the initial spike measured in a
484~ms time window is $(1.2 \pm 0.2) \times 10^{-7}$ erg cm$^{-2}$
s$^{-1}$. The hardness ratio between the 50--100 keV and the 25--50
keV bands for this initial spike is $1.2 \pm 0.3$, which is consistent
with the hardness of other \textit{Swift} SGRBs, though it is also
consistent with the LGRB population.  In Figure \ref{fig:harddur} we
plot the hardness and duration of GRB\ 080503 against other \Swift
bursts, resolving this burst and other short events with extended
emission separately into the spike and the extended tail. The
properties of the initial spike of GRB\ 080503 match those of the
initial spikes of other SGRBs with extended emission (and are
consistent with the population of short bursts lacking extended
emission), while the hardness and duration of the extended emission
are similar to that of this component in other short bursts.

\begin{figure}
\centerline{
\includegraphics[scale=0.6,angle=0]{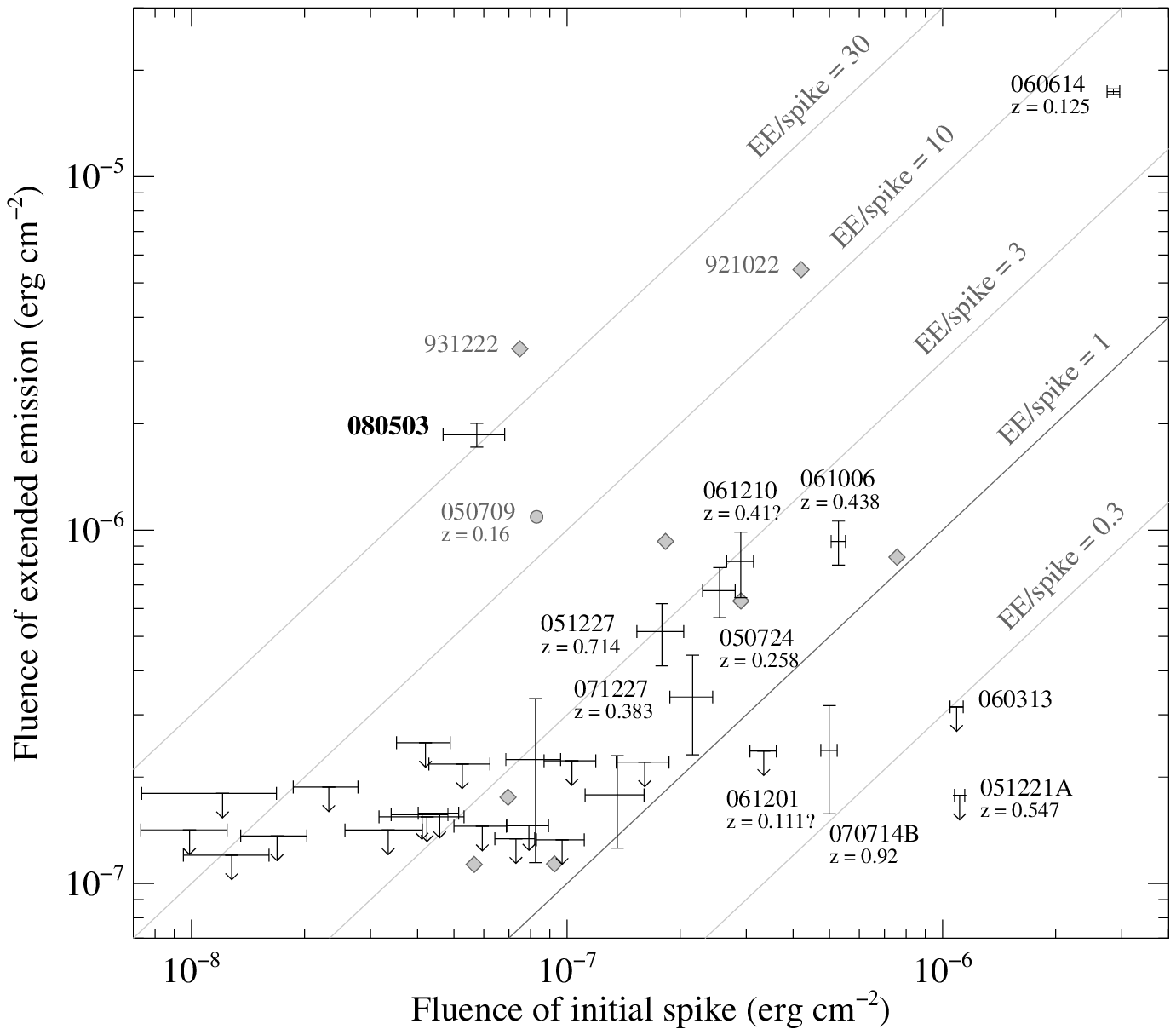}}
\caption{Fluences of the short initial spike versus the long
extended-emission episode for SGRBs and candidate SGRBs.  For \Swift
bursts this is measured in the 15--150 keV band.  For BATSE bursts
(diamonds) the values are calculated from the count rates in \citet{Norris+2006} and
fluences (20--100 keV) on the BATSE website.  HETE GRB\ 050709 (circle) is
taken from Table 2 of \citet{Villasenor+2005} and is in the 2--25 keV
band, which is significantly softer than the \Swift and BATSE
bandpasses.  In harder bandpasses the extended emission is likely to be
much fainter; this point should therefore be treated as an upper
limit.  BATSE and HETE short bursts without extended emission are not
shown.  Several properties are worthy of note.  First, the
extended-to-prompt ratio shows large variance, quite unlike the
observed T90 values and hardness ratios.  Second, the large majority
of \Swift events without extended emission are very faint bursts ---
the limits on the extended counterpart are not strongly constraining,
although strongly extended emission-dominated events like GRB\ 080503
do appear to be rare.  Third, events with bright extended
emission have a wide range of short-spike fluence; the two
values are not correlated.  Events with known redshift are labeled; no
clear trends with distance are evident.}
\label{fig:fluenceratio}
\end{figure}

The fluence of the extended emission measured from 5~s to 140~s after
the BAT trigger in the 15--150 keV bandpass is $(1.86 \pm 0.14) \times
10^{-6}$ erg cm$^{-2}$.  The ratio of this value to the spike fluence
is very large ($\sim$30 in the 15--150 keV band), higher than that of
any previous \Swift short (or possibly short) event including GRB\
060614.  It is not, however, outside the range measured for BATSE
members of this class, which
have measured count ratios up to $\sim$40 \citep[GRB\ 931222,][]
{Norris+2006}.  In Figure \ref{fig:fluenceratio}, we plot the fluences
in the prompt versus extended emission of all \Swift SGRBs to date.
BATSE bursts are overplotted as solid gray triangles; HETE event
GRB\ 050709 is shown as a star.  The two properties appear essentially
uncorrelated, and the ratio has a wide dispersion in both directions.
Although only two \Swift events populate the high extended-to-spike
ratio portion of the diagram (and the classification of GRB\ 060614 is
controversial), the difference in this ratio between these and more
typical events is only about a factor of 10, and the
intermediate region is populated by events from BATSE and
HETE\footnote{However, the HETE fluence ratio is in a very different
bandpass, and the actual ratio may be significantly lower than the
plotted ratio}, suggesting a continuum in this ratio across what are
otherwise similar events.

Lag analysis \citep{Norris+2000} has also been used as a short-long
diagnostic.  For GRB\ 080503, the spectral lag between the 50--100~keV 
and the 25--50~keV bands using the light curves in the 16 ms binning is $1 \pm 15$ ms
(1$\sigma$ error), consistent with zero and characteristic of
short-hard GRBs.  Unfortunately, the signal is too weak to measure the
spectral lag for the extended emission which dominates the fluence.
While lag can vary between pulses in a GRB \citep{Hakkila+2008} and
short pulses typically have short lags, even very short pulses 
in canonical long GRBs have been observed to have non-negligible 
lags \citep{Norris+2006}.

Based on all of these arguments, we associate GRB\ 080503 with the
``short'' (Type I) class.  Regardless of classification, however, the
extremely faint afterglow of this burst appears to
be a unique feature.  In fact, as we will show, while the extremely
low afterglow flux is more reminiscent of SGRBs than LGRBs, relative
to the gamma-rays the afterglow is so faint that this event appears
quite unlike any other well-studied member of either population to
date.

\subsection{UVOT Observations}
\label{sec:uvot}

The \Swift UV-Optical Telescope (UVOT) began observations of the field
of GRB~080503 at 83~s after the trigger, starting with a finding chart
exposure in the White filter at $t = 85$--184~s.  No source is
detected within the XRT position to a limiting magnitude of $>$20.0
\citep{GCN7675}.  A sequence of filtered observations followed, and
then additional White-band exposures.  The transient is not detected in any
exposure.  Because of the deep Gemini data shortly thereafter, these
additional limits do not constrain the behavior of the optical counterpart
and are not reported or reanalyzed here.  A summary of the
the subsequent UVOT observations is given by \cite{GCN7675}.
 
\subsection{Keck Observations}
\label{sec:keck}

Shortly after the GRB trigger we slewed with the 10~m Keck-I
telescope (equipped with LRIS) to the GRB position.
After a spectroscopic integration on a point source near the XRT position
that turned out in later analysis to be a faint star, we acquired 
(between 13:38:37 and 13:57:02)
imaging in the $B$ and $R$ filters simultaneously.  Unfortunately,
because the instrument had not been focused in imaging mode prior to
the target of opportunity, these images are of poorer quality
and less constraining than Gemini images (see below) taken at similar
times. The optical transient (OT) is not detected in either filter.
Magnitudes (calibrated using the Gemini-based calibration,
\S\ref{sec:gemini}) are reported in Table \ref{tab:photometry}.

On May 8 we used long-slit ($1''$ wide)
spectroscopy with LRIS \citep{Oke+1995} on Keck I to obtain spectra
of two relatively bright galaxies 13\arcsec\ SE of the afterglow
position.  We calibrated the two-dimensional spectra with standard arc
and internal flat exposures.  We employed the 600 line mm$^{-1}$ grism
(blue camera) and 600 line mm$^{-1}$ grating blazed at 10,000~\AA\
(red camera).  The data were processed with the
LowRedux\footnote{http://www.ucolick.org/$\sim$xavier/LowRedux/index.html;
developed by J. Hennawi, S. Burles, and J. X. Prochaska.} package
within
XIDL\footnote{http://www.ucolick.org/$\sim$xavier/IDL/index.html .}.
Both objects show the same emission lines, at common observed
wavelengths of $\lambda_{\rm obs} \approx$ 5821, 6778.8, 7592.2,
7745.6, and 7820 \AA.  The latter two are associated with the H$\beta$
and [O~III] $\lambda$5007 lines, respectively, identifying this system
to be at $z = 0.561$.

While the placement of the slit in the target-of-opportunity 
spectroscopic on May 3 did not cover the location of the transient, 
a third, serendipitous object along the slit shows a
single emission line at $\lambda_{\rm obs} \approx 6802.9$~\AA\ and a
red continuum. We tentatively identify this feature as unresolved
[O~II] $\lambda$3727 emission and estimate its redshift to be 0.8245.
This source is far (31$''$) from the OT position, at
$\alpha$~=~19\fh06\fm31\fs.1,
$\delta$~=~+68\arcdeg48\arcmin04\arcsec.3.


\subsection{Gemini Observations}
\label{sec:gemini}

We also initiated a series of imaging exposures using GMOS on the Gemini-North
telescope.  The first image was a single 180~s $r$-band exposure,
beginning at 13:24, 58 min after the \Swift trigger.  We then cycled
through the $g$, $r$, $i$, and $z$ filters with $5 \times 180$~s per
filter. A second $g$ epoch was subsequently attempted, but the images
are shallow due to rapidly rising twilight sky brightness.

The following night (May 4) we requested a second, longer series of images
at the same position.  Unexpectedly, the transient had
actually brightened during the intervening 24 hr, so we continued to
observe the source for several additional epochs.  
The next night (May 5), we acquired $r$-band images ($9 \times
180$~s), followed by a long nod-and-shuffle spectroscopic integration,
and concluded with $4 \times 180$~s exposures in each of the $g$ and
$i$ bands.  On May 6 and 7, we acquired long $r$-band imaging only
($14 \times 180$~s on May 6 and $16 \times 180$~s on May 7).  Finally,
on May 8, we acquired a long $K$-band integration using NIRI, nearly simultaneous
with the {\it HST} observations (\S \ref{sec:hst}) at the same epoch.

Optical imaging was reduced using standard techniques via the Gemini
IRAF package\footnote{IRAF is distributed by the National Optical
Astronomy Observatory, which is operated by the Association of
Universities for Research in Astronomy (AURA) under cooperative
agreement with the National Science Foundation.}.  Magnitudes were
calculated using seeing-matched aperture photometry and calibrated
using secondary standards.  The standard star field SA~110 was observed on 
the nights of May~3, May~4, May~5, and May~8;  catalog magnitudes 
\citep{Landolt1992} were converted to $griz$ using the equations from
\cite{Jester+2005} and used to calibrate 23 stars close to the GRB
position (Table \ref{tab:calibstars}).  

In an attempt to measure or constrain the redshift of GRB~080503, we 
obtained a nod-and-shuffle long-slit spectroscopic integration of the 
positions of the optical transient and the nearby faint galaxy S1
(Figure \ref{fig:image3}).  Two exposures of 1320~s each were obtained
starting at 12:20 on 2008 May 05.
Unfortunately, even after sky subtraction and binning, no clear trace
is observed at the position, and no line signatures are apparent.  The
redshift of the event is therefore unconstrained, except by the
$g$-band photometric detection which imposes a limit of approximately
$z < 4$.

We began near-infrared observations of GRB~080503 on 2008 May 08 at
12:46, roughly simultaneous with the {\it HST} measurement (\S
\ref{sec:hst}).  All images were taken in the $K_s$ band with NIRI. We
employed the standard Gemini-N dither pattern for each of the 30~s
exposures.  In all, 92 images were taken yielding a total time on
target of $\sim$1.5 hr. The data were reduced and the individual
frames were combined in the usual way using the ``gemini'' package
within IRAF.  There is no detection of a source at the location of the
optical transient.  The nearby faint galaxies (S1 and S4) are also
undetected.  Calibrating relative to the 2MASS catalog (excluding
stars near the edge of the image because NIRI is known to suffer from
fringing), we derive an upper limit of $K_s >$ 22.47 mag ($3\sigma$).

\begin{figure}
\centerline{
\includegraphics[scale=0.7,angle=0]{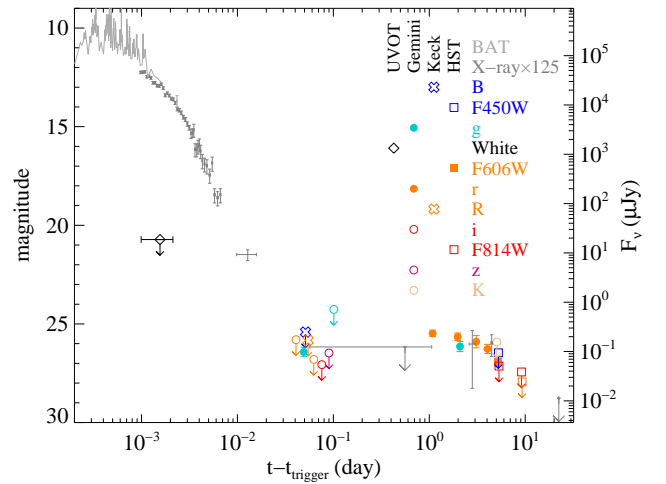}}
\caption{X-ray and optical light curves of GRB 080503.  The optical
bands have been shifted to the $r$ band assuming an optical spectral
index of $\beta = 1.2$; the X-ray light curve has been shifted by a
factor of 125 to match the optical (corresponding to $\beta_{OX} =
0.75$).  The BAT light curve is extrapolated into the X-ray band using
the high-energy spectrum.  3$\sigma$ upper limits are shown with arrows.}
\label{fig:lcurve}
\end{figure}

All optical photometry, in conjunction with the space-based measurements
from \Swift \ and Chandra, is plotted in Figure \ref{fig:lcurve}.

\subsection{Hubble Space Telescope Observations}
\label{sec:hst}

\begin{figure*}
\centerline{
\includegraphics[scale=0.8,angle=0]{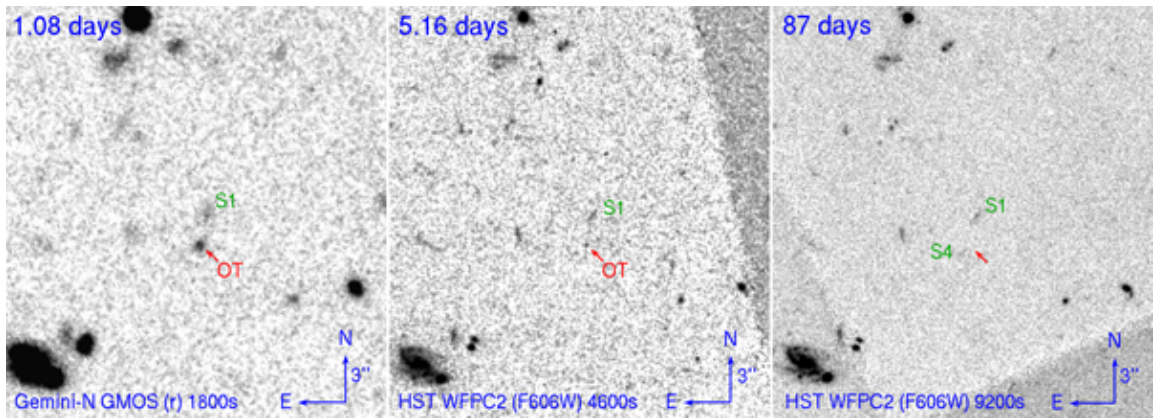}}
\caption{Ground-based and space-based images showing the evolution of
the faint OT associated with GRB 080503.  The transient peaked at
about $t = 1$~d, shown in an image from Gemini-North at left.
Thereafter it faded rapidly and is barely detected in the first {\it
HST} epoch in F606W only.  Later observations failed to reveal a
galaxy coincident with the transient position.  Two very faint nearby
(but non-coincident) galaxies are designated ``S1'' and ``S4.''}
\label{fig:image3}
\end{figure*}

Given the unusual nature of the afterglow, and the indications of a
Li-Paczy\'{n}ski-like light curve in the first two days, we
proposed\footnote{Program GO-DD 11551; PI Bloom.} to observe the field
of GRB\ 080503 with the Wide-Field Planetary Camera (WFPC2) on {\it HST}.  
Filter changes, depth, and cadences
were chosen to confirm or refute the basic predictions of the
\citet{LiPaczynski1998} model (see Fig. \ref{fig:minisn} and \S
\ref{sec:minisn}). The localization region was observed in three epochs
on 2008 May 8, May 12, and July 29.  A set of F450W (1 orbit),
F606W (2 orbits), and F814W (1 orbit) observations were obtained during
the first visit, with F606W (2 orbits) and F814W (2 orbits) in the
second visit, and finally a deep (4 orbit) observation in F606W 
in the third visit. Observations were dithered (a 3-point
line dither for the first epoch of F450W and F814W, and a standard 
4-point box for all other observations).  The data were reduced in the
standard fashion via {\tt multidrizzle}, while the pixel scale was
retained at the native $\sim 0.1$\arcsec pixel$^{-1}$.

At the location of the afterglow in our first-epoch F606W image we
found a faint point source, with a magnitude of F606W = 27.01 $\pm$
0.20 after charge-transfer efficiency correction following \cite{Dolphin2000}.  
Our other observations show
no hint of any emission from the afterglow or any host galaxy directly
at its position. We derived limits on any object at the GRB position
based on the scatter in a large number ($\sim 100$) of blank apertures
placed randomly in the region of GRB 080503. The limits for each frame
are shown in Table~\ref{tab:photometry}. In addition, a stacked frame of
all our F814W observations yields F814W $> 27.3$ mag. A combination of
all but our first-epoch F606W observations provides our deepest limit
of F606W $> 28.5$ mag (3$\sigma$), in a stacked image with exposure
time 13,200~s.  Therefore any host galaxy underlying GRB\ 080503 must
be fainter than that reported for any other short burst.

Although there is no galaxy directly at the GRB position, there are
faint galaxies close to this position which are plausible hosts. In
particular, our stacked image of all the F606W observations shows a
faint galaxy $\sim 0.8$\arcsec \ from the afterglow position, with
F606W(AB) = 27.3 $\pm$ 0.2 mag (designated ``S4'' in Figure
\ref{fig:image3}).  Although faint, this galaxy is clearly extended,
with its stellar field continuing to $\sim$0.3\arcsec \ from the GRB
position.  (It is plausible that deeper observations or images in
redder wavebands may extend its disk further, but we have no evidence
that this is the case.)  Additionally, there is a brighter galaxy
(``S1,'' F606W $\approx 26.3$ mag) $\sim 2''$ to the north of the
afterglow position, also visible in the Gemini images.  Given the
faintness of these galaxies and the moderate offset from the afterglow
position, the probability of chance alignment is nontrivial (a few
percent, following \citealt{Bloom+2002}), and we cannot make firm
statements about their association with GRB\ 080503.

\begin{figure}
\centerline{
\includegraphics[scale=0.35,angle=0]{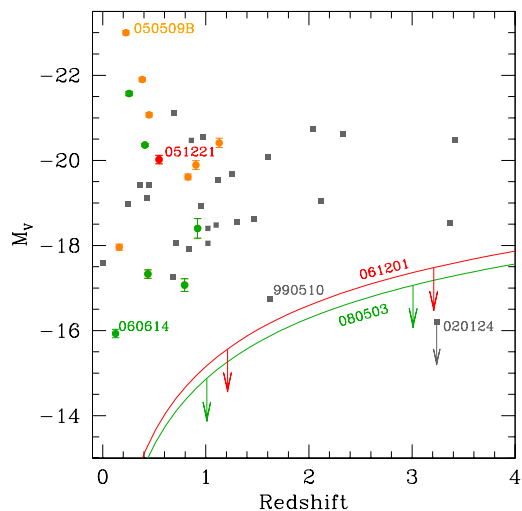}}
\caption{The absolute magnitudes and redshifts for a sample of both long (grey squares, from \citealt{Fruchter+2006}) and short GRB hosts.  Bursts with extended emission are marked in green and bursts without extended emission are red; orange denotes SGRBs too faint for a strong limit on extended emission fluence to be inferred.  The two solid lines represent ``host-less'' SGRBs 061201 and 080503, and are extrapolated based on the observed limits. Due to the poor wavelength sampling of many faint GRB hosts the absolute magnitudes have been obtained assuming a flat spectrum K-correction $M_V = V - DM + 2.5 \log(1+z)$, where $DM$ is the distance modulus. We have assumed a $\Lambda$CDM cosmology with $\Omega_M = 0.27$, $\Omega_{\Lambda} = 0.73$ and $H_0 = 72$ km s$^{-1}$ Mpc$^{-1}$.  The nondetection of a host for GRB 080503 implies either that it lies at higher redshift than the majority of the SGRB population, that it originates from a host which is much fainter than the median, or that it has been ejected to a sufficient distance from its host that it can no longer be firmly associated with it. Such deep limits to hosts underlying GRBs are rare, with only a single LGRB (020124, \citealt{Berger+2002}) undetected in deep HST imaging (out of a sample of $\sim 50$), while two SGRBs (of roughly 15 with good optical positions) are undetected to similar limits.}
\label{fig:hostplot}
\end{figure}

The extremely deep limit on a host galaxy puts GRB~080503 in very rare
company.  Among short bursts, no comparably deep limit exists for any
previous event except GRB\ 061201, although a study with deep {\it HST} imaging of
short-burst hosts has yet to be published.  However, ground-based searches for
hosts of other SGRBs with subarcsecond positions have identified
coincident host galaxies in 9 of 11 cases.  The two exceptions are
GRB~061201 \citep{Stratta+2007} and GRB~070809 \citep{GCN7889}; both
of these appear at relatively small physical offset from nearby
spirals which have been claimed as host candidates.  Short GRB\ 070707
has a coincident host with $R=27.3$ mag \citep{Piranomonte+2008}, about the same as the
magnitude of the nearest galaxy to the GRB\ 080503 OT position.  In
fact, even compared to long bursts, the lack of host galaxy is unusual; only
five events have host-galaxy measurements or limits fainter than 28.5
mag.

There are two general possibilities to explain this extreme faintness.
First, GRB\ 080503 could be at high redshift ($z > 3$), or at
moderately high redshift in a very underluminous galaxy (at $z \approx
1$, comparable to the highest-$z$ SGRBs detected to date, $M_B < -15$
mag).\footnote{GRB\ 080503 could also be at moderate redshift $z=1-3$ in
a moderately large but extremely dusty galaxy.  Even then, our $K$ 
nondetection imposes strong constraints on the size of the object,
and the relatively blue $g-r$ afterglow color suggests that the 
environment of the GRB is not particularly dust-obscured.}
A bright ``short'' GRB at very high redshift would impose a
much larger upper end of the luminosity distribution of these events
than is currently suspected.  An extremely underluminous host would 
also be surprising under a model associating SGRBs with old stars, 
since the bulk of the stellar mass at moderate redshifts is still
in relatively large galaxies \citep{Faber+2007}. 

Second, GRB\ 080503 could be at low redshift but ejected a long
distance from its host.  To further examine this possibility, we have
estimated the probabilities \citep[following][]{Bloom+2002} of a
statistically significant association with other bright galaxies in
the field.  A rather faint spiral galaxy is located 13\arcsec\ SE of
the afterglow position (J2000 coordinates $\alpha =
19\fh06\fm31\fs.7$, $\delta = 68\arcdeg47\arcmin27\arcsec.9$; visible
in the bottom-left corner of Figure \ref{fig:image3}) and has $r =
21.7$ mag and $z = 0.561$ (\S \ref{sec:keck}).  The probability
that this is a coincidence is of order unity.  We also searched NED
and DSS image plates for very bright nearby galaxies outside the
field.  The nearby ($D \approx 5$ Mpc) dwarf galaxy UGC 11411 is
located at an offset of 1.5$^\circ$; again the chance of random
association is of order unity.  There are no other nearby galaxies of
note.  While a low probability of random association does not rule out
an association with one of these objects (a progenitor that escapes
its host-galaxy potential well and has a sufficiently long merger time
will be almost impossible to associate with its true host), it
prevents us from making an association with any confidence.

\subsection{Swift XRT analysis}
\label{sec:xrt}

The \Swift X-ray telescope began observing GRB~080503 starting
$\sim$82~s after the burst, detecting a bright X-ray
counterpart.  Observations continued during the following hour and in
several return visits.

The XRT data were reduced by procedures described by
\citet{Butler+2007a}.  The X-ray light curve, scaled to match the
optical at late times, is shown in Figure \ref{fig:lcurve}.  Despite
the bright early afterglow, the flux declined precipitously and no
significant signal is detected during the second through fourth
orbits.  A marginally significant detection is, however, achieved
during a longer integration a day later.

\begin{figure}
\centerline{
\includegraphics[scale=0.5,angle=270]{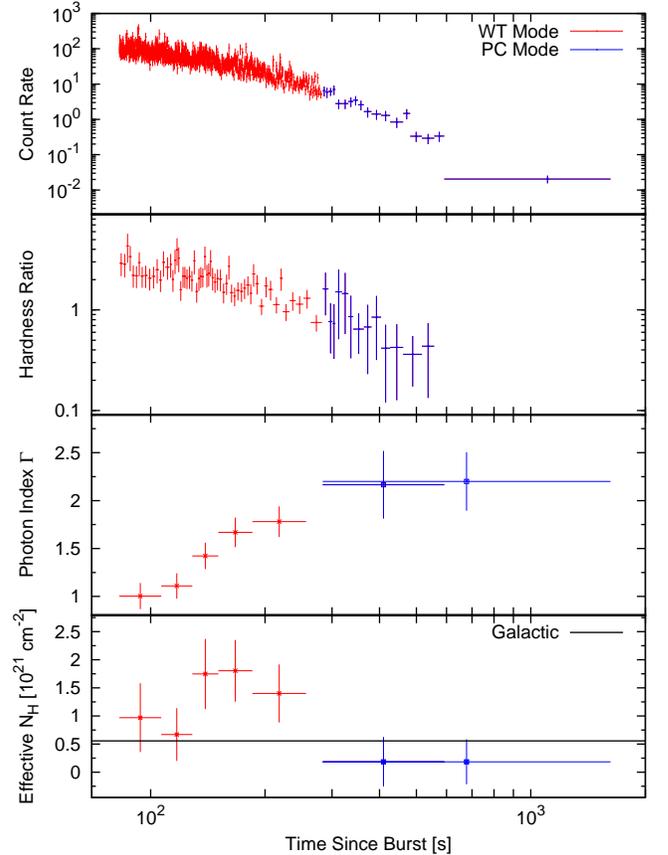}}
\caption{(a) The 0.3--10.0 keV X-ray flux measured by the XRT declines
rapidly following the bursts. (b) The ratio of counts in the 1.3--10.0
keV to 0.3--1.3 keV bands also declines.  (c,d) The spectrum is well
modelled by an absorbed power law, although the effective column
density $N_H$ appears to unphysically rise and decline during the
observations (see text).}
\label{fig:xrt}
\end{figure}

The X-ray hardness ratio decreases, as does the 0.3--10.0 keV count
rate, during the course of the early observations (Figure
\ref{fig:xrt}a,b).  Absorbed power-law fits to the evolving spectrum
are statistically acceptable ($\chi^2/$dof $\approx$ 1) and yield a
photon index $\Gamma$ which increases smoothly with time and an
H-equivalent column density $N_H$ that apparently rises and then falls
in time (Figure \ref{fig:xrt}c,d).  This unphysical $N_H$ variation is
commonly observed in power-law fits to the XRT emission following BAT
GRBs and XRT flares \citep[see, e.g.,][]{Butler+2007b}; it suggests
that the intrinsic spectrum, plotted on a log-log scale, has
time-dependent curvature.  In fact, we find that the combined BAT and
XRT data are well fit by a GRB model \citep{Band+1993} with constant
high- and low-energy photon indices and a time-decreasing break energy
that passes through the XRT band during the observation.

The amount of physical column density that contributes to the
effective $N_H$ in Figure \ref{fig:xrt}c can be estimated at early or
late times, when the effective $N_H$ is near its minimum, or from the
\cite{Band+1993} GRB model fits. We find $N_H = 5.5^{+1.5}_{-0.9} \times 10^{20}$ cm$^{-2}$,
comparable to the Galactic value of $N_H = 5.6 \times 10^{20}$
cm$^{-2}$, indicating that the host-galaxy hydrogen column is minimal.

\subsection{Chandra X-Ray Observatory Observations}
\label{sec:chandra}

Under Director's Discretionary Proposals 09508297 and 09508298, we
conducted imaging using the Chandra X-Ray Observatory ACIS-S on two
occasions.  During the first integration (2008-05-07 19:18:23 to
2008-05-08 04:09:59) an X-ray source is detected at $\alpha$ =
19\fh06\fm28\fs.76, $\delta$ = +68\arcdeg47\arcmin35\arcsec.3 (J2000,
0.5\arcsec \ uncertainty), consistent with the position of the optical
afterglow.  This source was not detected during the second epoch
(2008-05-25 18:11:36 to 2008-05-26 03:04:28), limiting the decay rate
to steeper than approximately $t^{-1.6}$.

Minimizing the \cite{Cash1976} statistic, we find the Chandra spectrum
to be acceptably fit by an absorbed power law with $\beta = 0.5 \pm
0.5$ and unabsorbed flux $F_X = (1.5 \pm 0.7) \times 10^{-14}$ ergs
cm$^{-2}$ s$^{-1}$ (0.3--10~keV).  We assume Galactic absorption only.

We attempted to use the photon arrival times to constrain the temporal
index ($\alpha$) assuming power-law brightening or fading behavior
\citep{Butler+2005}.  The exposure time is short compared to the time
elapsed since the GRB, precluding strong constraints.  Although the
data do marginally favor brightening behavior ($\alpha = -13 \pm 7$), in
contrast to the well-established optical fading at this point, we do
not consider this to be a strong conclusion.

\section{Modeling and Interpretation}
\label{sec:model}

\subsection{The Origin of the Rapid Decay Phase}
\label{sec:hilatitude}

\begin{figure}
\centerline{
\includegraphics[scale=0.45,angle=0]{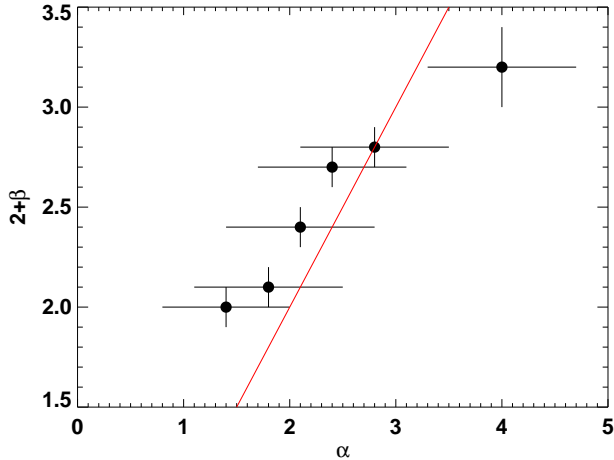}}
\caption{Decay index $\alpha$ versus spectral index $\beta$ (+2)
during the rapid-decay phase of the external power-law.  For a purely
power-law spectrum a closure relation $\alpha$ = 2 + $\beta$ is
predicted by the high-latitude (curvature) model; this is
approximately obeyed as shown by the solid line.  For more complicated
spectra this relation may not be obeyed exactly.}
\label{fig:alphabeta}
\end{figure}

Immediately after the prompt emission subsides, the X-ray light curve
(Fig. \ref{fig:alphabeta}) is observed to decline extremely rapidly
($\alpha = 2$--4, where $\alpha$ is defined by $F_{\nu} \propto
t^{-\alpha}$), plummeting from a relatively bright early X-ray flux to
below the XRT detection threshold during the first orbit.  Although a similar
rapid early decline is seen in nearly all GRBs for which
early-time X-ray data are available \citep{O'Brien+2006}, GRB\ 080503
probably constitutes the most dramatic example of this on record: the
decline of $\sim$6.5 orders of magnitude from the peak BAT flux is
larger by a factor of $\sim$ 100 than observed for the reportedly 
``naked'' GRB~050421 \citep{Godet+2006} and comparable to the decline
of two other potentially naked \Swift events described by
\cite{Vetere+2008}.  The lack of contamination of this phase of the
GRB by any other signature (X-ray flares or a standard afterglow) affords an 
excellent test for models of this decay component.

An afterglow interpretation can be ruled out almost immediately.  In
addition to the difficulties faced by such a model in explaining the
very sharp decay index, continuous spectral
softening, and smooth connection with the prompt emission (all of
which are commonly observed in the rapid decay phase of other GRBs),
the early UVOT White
measurement ($\lesssim$220 $\mu$Jy at 85--184 s) imposes a limit
on the X-ray to optical spectral slope of $\beta_{\rm OX} < -0.5$
(using the convention $F_\nu \propto \nu^{-\beta}$) that is very difficult
to explain as afterglow emission, but is consistent with the
low-energy tail of prompt-emission spectra.

While the origin of the rapid-decay phase observed in most X-ray light curves
is still not settled, the most popular interpretation is high-latitude 
emission \citep{Kumar+2000}, also referred to as the curvature effect.
In this scenario, after the prompt emission ends some photons still
reach us from increasingly larger angles relative to the line of sight
(to the central source) due to a longer path length induced by the
curvature of the (usually assumed to be quasi-spherical) emitting
region (or shell). Such late photons correspond to a smaller Doppler
factor, resulting in a relation between the temporal and spectral
indexes, $\alpha = 2 + \beta$, that holds at late times ($t-t_0 \gg
\Delta t$) for each pulse in the prompt light curve (of typical width 
$\Delta t$ and onset time $t_0$) where $\beta = -d\log F_\nu/d\log\nu$
and $\alpha = -d\log F_\nu/d\log(t-t_0)$. The total tail of the prompt
emission is the sum of the contributions from the different pulses. At
the onset of the rapid-decay phase the flux is usually dominated by
the tail of the last spike in the light curve, and therefore can
potentially be reasonably fit using a simple single-pulse model with
$t_0$ set to near the onset of this last spike. At later times the
tails of earlier pulses can become dominant. At sufficiently late
times both $t-t_0 \gg \Delta t$ and $t \gg t_0$ (i.e., $t - t_0 \approx
t$) for all pulses, and the relation $\alpha = 2+\beta$ is reached for
$t_0 = 0$ (i.e., setting the reference time $t_0$ to the GRB trigger
time).  In GRB~080503 the large dynamic range enables us to probe
this late regime; as shown in Figure \ref{fig:alphabeta},
which displays $\alpha$ versus $2+\beta$ for the rapid-decay phase
using $t_0 = 0$, the relation $\alpha = 2+\beta$ roughly holds, as
expected for high-latitude emission.  

While the above discussion suggests that high-latitude emission is a
viable mechanism for the rapid-decay phase in GRB~080503, a more
careful analysis is called for, especially since assuming an intrinsic
power-law spectrum during the rapid-decay phase requires an unphysical
time-variable $N_H$; a better and more physical description is
provided by using a fixed Galactic value for $N_{\rm H}$ and an
intrinsic \cite{Band+1993} spectrum whose peak energy passes through the XRT range
(see \S \ref{sec:xrt}).  A more detailed analysis of this event (and
others) in the context of the high-latitude model and possible
alternatives using this model will be forthcoming in future work.

\subsection{Constraining the External Density from Lack of Early 
Afterglow Emission}
\label{sec:environs}

\begin{figure*}
\centerline{
\includegraphics[scale=1.0,angle=0]{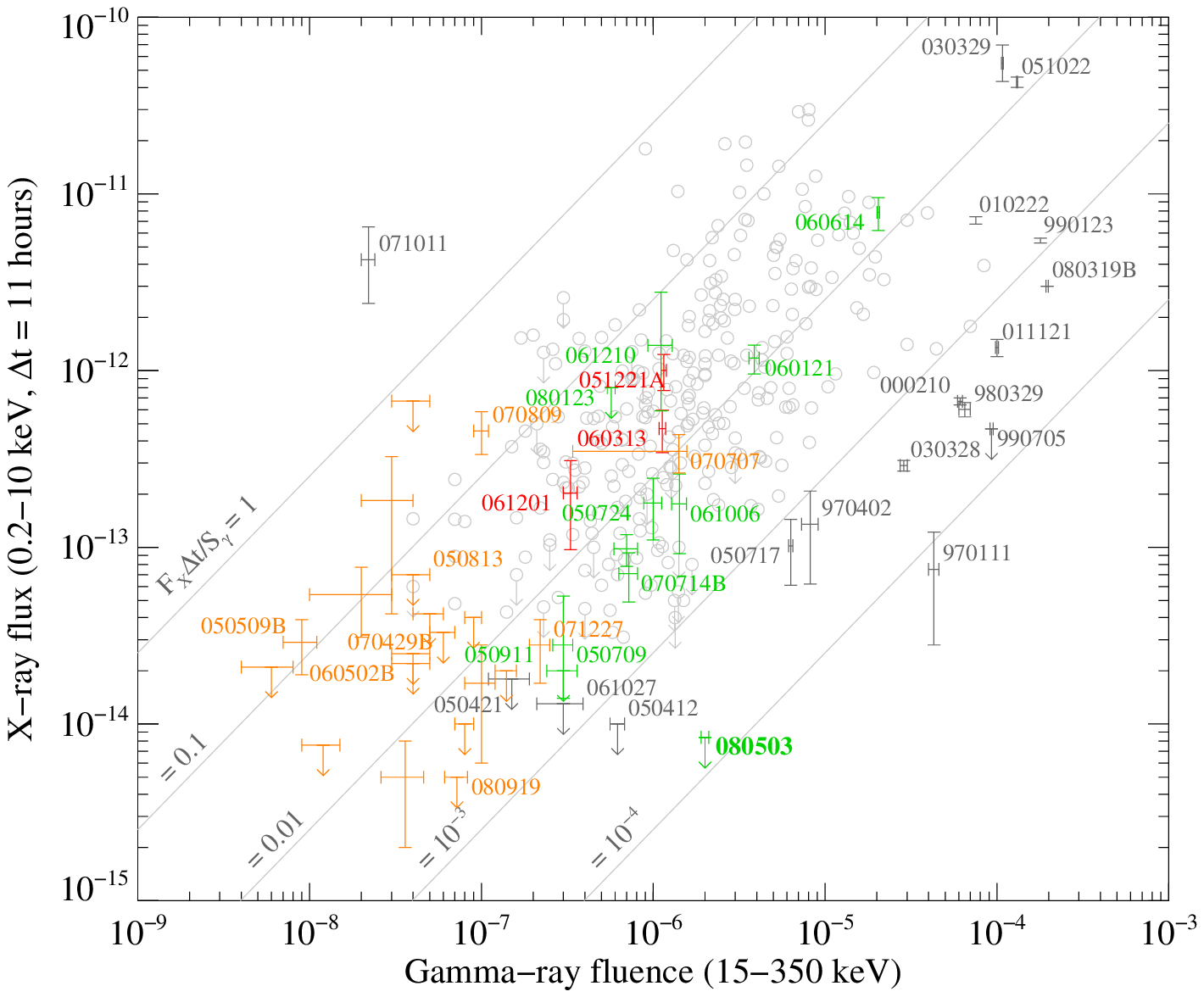}}
\caption{Comparison of the total gamma-ray fluence (15--150 keV)
versus X-ray flux (0.2--10 keV) at 11 hr post-burst for all GRBs with
X-ray afterglows, based on Figure 4 and Tables 1--2 of
\citet{Nysewander+2008} supplemented with our own re-evaluation of the
upper limits on events without detections after $\sim$10 hr using
the \Swift XRT repository \citep{Evans+2007} and other primary references
listed in \cite{Nysewander+2008}.  New SGRBs in 2008 have been added, 
along with the extremely bright GRB 080319B \citep{Bloom+2008}.  
Long bursts are shown in gray, short bursts without extended emission 
in red, faint short bursts with poor constraints on extended emission 
in orange (as in Figure \ref{fig:harddur}), and
short bursts with extended emission (including the ambiguous GRB\
060614) in green.  Prominent events are labeled.  Almost all events
with detections fall along an approximately linear relation indicating
a roughly constant prompt-to-afterglow ratio; most upper limits are
not inconsistent with this.  GRB\ 080503 (plotted as an upper limit,
though the detection by Chandra at several days after trigger suggests
that the flux cannot be much less than this) is strongly discrepant
compared to nearly all previous events.  GRB\ 970111 is the first
burst for which rapid X-ray observations were conducted and its
general faintness appears to be real \citep{Feroci+1998}; however,
based on the plot in the supplementary material of
\cite{dePasquale+2006} the afterglow flux at 11~hr may be somewhat
underestimated.}
\label{fig:nyse}
\end{figure*}

The faintness of the early afterglow is very striking.  Any afterglow 
emission for this event was unlikely to be brighter than about 
$\sim 1$ $\mu$Jy at optical wavelengths and $10^{-2}$ $\mu$Jy in X-rays
at any time after about 1 hr (and  if the late afterglow peak were due 
to a non-afterglow signature, a possibility we consider in 
\S \ref{sec:minisn}, these limits would be even more stringent.)  
Our early optical limits are the deepest for any GRB on record at this epoch
\citep{Kann+2008}.  If the observed emission at $t > 1$~d is due to a
mini-SN or other process, the absence of an afterglow is even more
notable.  Figure \ref{fig:nyse} shows the X-ray flux at
$11\;$hr, $F_X(11\,{\rm hr})$, and the fluence of the prompt
$\gamma$-ray emission, $S_\gamma$, for GRB~080503 together with a
large sample of both LGRBs and SGRBs (data taken from Figure 4 of
\citealt{Nysewander+2008}, but modified slightly as described in the
caption.)  GRB\,080503 immediately stands out as a dramatic outlier,
with an $F_X/S_{\gamma}$ several orders of magnitude below that of
the general population, indicating a poor conversion of the energy left 
in the flow after the prompt gamma-ray emission into afterglow 
(emission from the external forward shock).  A natural explanation 
for this difference is a very low external density.

Using the upper limit on the X-ray flux, $F_X(11\,{\rm hr}) <
8.4\times 10^{-15}\;{\rm erg\;cm^{-2}\;s^{-1}}$, and the measured
fluence, $S_\gamma = (1.7 \pm 0.1)\times 10^{-6}\;{\rm erg\;cm^{-2}}$,
we derive constraints on the external density, $n = n_0\;{\rm
cm^{-3}}$.  Following \citet{GKP06}, it is convenient to use the X-ray
afterglow efficiency, $\epsilon_X(t) \equiv t L_X(t)/E_{\rm
k,iso}(t)$. We can relate the isotropic equivalent kinetic energy in
the afterglow shock, $E_{\rm k,iso}$, to the measured fluence by using
the ratio $\eta_{\rm k\gamma} \equiv E_{\rm k,iso}/E_{\rm\gamma,iso}$,
which is expected to be of order unity. This gives
\begin{equation}\label{eps_X_obs}
\epsilon_X = \frac{t F_X(t)}{\eta_{\rm k\gamma}S_\gamma},\quad\quad
\epsilon_X(t = 11\,{\rm hr}) < 8.0\times 10^{-5}\eta_{\rm k\gamma}^{-1}\ ,
\end{equation}
where $L_X(t)$ in the definition of $\epsilon_X$ is interpreted here
as evaluated at $t = 11$~hr and an energy range of 2--10 keV
(converted from our reported 0.3--10 keV value assuming $\beta \approx
-1$) in the observer frame. This makes it easier to compare this value
to the one derived from standard afterglow theory, as is done next.

The value of $\beta_{\rm OX} \approx 0.7$ suggests $p \approx 2.4$ if
the cooling break frequency is above the X-rays, $\nu_c > \nu_X$, and
a smaller value of $p$ if $\nu_c < \nu_X$.  If $\nu_c < \nu_X$ then
for $p \approx 2.2$ and $\epsilon_B \ll \epsilon_e$ \citep[using
eq.~7 of][]{GKP06},
\begin{eqnarray}\label{eps_X_AG}
\epsilon_X(t = 11\,{\rm hr}; \nu_c < \nu_X) & \approx &
10^{-3}\epsilon_{e,-1}^{p-3/2}\epsilon_{B,-2}^{p/4}E_{\rm k,iso,52}^{(p-2)/4} \\
& \sim & 10^{-3}\epsilon_{e,-1}^{0.7}\epsilon_{B,-2}^{0.55}E_{\rm k,iso,52}^{0.05}\ ,
\end{eqnarray}
where $\epsilon_e = 0.1\epsilon_{e,-1}$, $\epsilon_B =
0.01\epsilon_{B,-2}$, and $E_{\rm k,iso} = 10^{52}E_{\rm
  k,iso,52}\;$erg.  There is no dependence on the external density as
long as $\nu_c < \nu_X$, and the dependence on $E_{\rm k,iso}$ is
extremely weak.  It does have some dependence on $\epsilon_e$ and
$\epsilon_B$. However, reproducing the value derived in
eq.~(\ref{eps_X_obs}) requires these shock microphysical parameters to
assume very low values -- not out of the question but on the low end
of the values inferred from modeling of the best-monitored GRB
afterglows. This is assuming a reasonable efficiency of the gamma-ray
emission, $\epsilon_\gamma \lesssim 0.5$, leaving at least a
comparable kinetic energy in the outflow that was transferred to the
shocked external medium before $11\;$hr, 
$\eta_{\rm k\gamma} \approx (1-\epsilon_\gamma)/\epsilon_\gamma \gtrsim 1$. 
For typical values of
the shock microphysical parameters ($\epsilon_e \approx 0.1$ and
$\epsilon_B \approx 0.01$), eqs.~(\ref{eps_X_obs}) and
(\ref{eps_X_AG}) can be reconciled either if $\nu_c(11\,{\rm hr}) \gg
\nu_X$ (which as is shown below implies a very low external density),
or if $1 - \epsilon_\gamma \approx \eta_{\rm k\gamma} \ll 1$
(i.e., an extremely high gamma-ray efficiency that leaves very little
energy in the afterglow shock, compared to that emitted in
gamma-rays).

For a reasonable gamma-ray efficiency ($\epsilon_\gamma \lesssim 0.5$)
this suggests that $\nu_c > \nu_X$. In this case the value of
$\epsilon_X$ is reduced by a factor of $(\nu_c/\nu_X)^{1/2}$ compared
to its value for $\nu_c < \nu_X$ (that is given in
eq.~(\ref{eps_X_AG}) for $p \approx 2.2$) and is smaller by a factor of
$\sim$1.48 for $p \approx 2.4$ (that is inferred from the observed
value of $\beta_{\rm OX}$ for $\nu_c > \nu_X$). For a $\nu_X \approx 10^{18}\;$Hz 
(corresponding to $\sim 4\;$keV) this suggests
$\nu_c(11\,{\rm hr}) \gtrsim 10^{20}\;$Hz, which in turn
\citep[using the expression for $\nu_c$ from][]{GS02} implies
\begin{equation}
n \lesssim 5 \times 10^{-6}E_{\rm
k,iso,52}^{-1/2}\epsilon_{e,-1}^{-1}\epsilon_{B,-2}^{-1/2}\;{\rm cm^{-3}}\ .
\end{equation}
This dependence on the parameters is valid in the
limit of $\epsilon_B \ll \epsilon_e$, where $Y \approx
(\epsilon_e/\epsilon_B)^{1/2} \gg 1$ and $\nu_c \propto n^{-1}E_{\rm
k,iso}^{-1/2}(1+Y)^{-2}\epsilon_B^{-3/2} \propto n^{-1}E_{\rm
k,iso}^{-1/2}\epsilon_e^{-1}\epsilon_B^{-1/2}$. Therefore, the upper limit
on the external density cannot easily be increased by a large
factor.  This suggests a very low external density compared to typical
disk values ($n \approx 1$ cm$^{-3}$) or even a Galactic halo ($n \approx 10^{-3}$ cm$^{-3}$, \citealt{Basu})
but is of the same order as the intergalactic particle density 
($n \approx 10^{-6}$ cm$^{-3}$, \citealt{Hinshaw+2008}).  This result 
therefore provides strong evidence that this explosion occurred
far outside any galaxy.  (An intriguing alternative to this, however, would 
be if the burst occurred in a low-density pulsar cavity inflated by one of the
NSs in the precursor binary; \citealt{RosswogRR2003}.)

\subsection{Afterglow Models: Why the Delay?}
\label{sec:lateag}

The counterpart rebrightened during the second night of observations,
rising again above detectability in both the optical and X-ray bands.
The optical is far better constrained than the X-rays in this case:
the rise is at least 1.5 mag (a factor of $\sim 3$) and peaks between 0.1
and 2~d after the event, though most likely the peak is toward the end
of this period as the optical observations at 1--2~d are consistent
with constant flux.  Although the faint afterglow and sparse
observations preclude a careful search for chromatic behavior, the
X-ray emission shows a broadly similar temporal behavior as the
optical and is consistent with being on the same segment of a
power-law spectrum ($F_\nu \propto \nu^{-\beta}$), with a very
reasonable value of the optical to X-ray spectral slope for GRB
afterglows, $\beta_{\rm OX} \approx 0.7$. This suggests that they
arise from the same physical region, and probably also from the same
emission mechanism (most likely synchrotron emission from the forward
external shock, i.e. the afterglow; we will consider other models in
\S \ref{sec:minisn}).

A late peak ($t \approx 1$~d) is unusual for an afterglow but not
unprecedented.  Most such events are \emph{rebrightenings} and not
global maxima.  The most prominent examples of this have been long bursts, 
though some modest X-ray flaring has been observed in a few short GRBs
\citep{Fox+2005,Campana+2006}, and notably the
classification-challenged GRB\ 060614 had an optical peak
between 0.3--0.5 d.  Without deep imaging before our first Gemini exposure,
we cannot constrain the nature of an optical afterglow in the earliest
phases of GRB\ 080503.  However, it is clear that since this behavior
is consistent with that observed for at least some previous GRB afterglows, 
the observed light curve, like the SED, is consistent with an afterglow model.  
The cause of this delayed peak, however, remains an open question, which
we will now turn our attention to.

The similar temporal behavior of the X-ray and optical flux around
the observed peak argues against a
passage of a spectral break frequency (e.g., the typical synchrotron
frequency $\nu_m$ passing through the optical) as the source of the
late time peak in the light curve, and in favor of a hydrodynamic
origin. One possibility for such a hydrodynamic origin is the
deceleration time, $t_{\rm dec}$. However, such a late deceleration
time implies either an extremely low initial Lorentz factor of the
outflow, $\Gamma_0$, or an unreasonably low external density
\begin{eqnarray}
n_0 &\approx& \left[\frac{t_{\rm dec}}{42(1+z)\,{\rm s}}\right]^{-3} E_{\rm k,iso,51} 
\left(\frac{\Gamma_0}{100}\right)^{-8} \\ &\approx&
10^{-10}E_{\rm k,iso,51}
\left(\frac{\Gamma_0}{100}\right)^{-8} \\ &\approx& E_{\rm k,iso,51} 
\left(\frac{\Gamma_0}{5.7}\right)^{-8}\ 
\end{eqnarray}
\citep[see, e.g.,][]{Granot05,LR-RG05}, where we have
used $t_{\rm dec}/(1+z) \approx 1$~d.   

An initial Lorentz factor of $\Gamma_0 \gtrsim 100$ is typically required in order to overcome the compactness problem for the prompt GRB emission. This would in turn imply in our case an external density of $n \lesssim 10^{-10}\;{\rm cm^{-3}}$ that is unrealistically low, even for the the intergalactic medium (IGM). An external density typical of the IGM, $n_{\rm IGM} \sim 10^{-6}\;{\rm cm^{-3}}$ would require $\Gamma_0 \sim 30$.  This may or may not be a strong concern in this case: the constraints on the high-energy spectrum of the extended-emission component of short GRBs are not yet well-established\footnote{Note, however, that EGRET has detected high-energy emission including a $\sim 1\;$GeV photon \citep{Sommer+1994} in the extended prompt emission (lasting $\sim 50\;$s) of the short ($< 1\;$s) GRB~930131 \citep{Kouveliotou+1994}.}, and it is not yet certain that existing compactness constraints apply to this emission component, potentially allowing a lower minimum Lorentz factor than is required for SGRB initial spikes ({\it Fermi} has detected high energy emission up to $\sim 3\;$GeV from the short GRB~081024B, \citealt{GCN8407}) or for classical LGRBs.

An alternative hydrodynamic explanation for the late peak is if the
afterglow shock encounters a large and sharp increase in the external
density into which it is propagating. However, it would be very hard
to produce the required rise in the light curve up to the broad peak
due to a sudden jump in the external density \citep{NG07} unless a 
change in the micro-physical parameters accompanies the sharp density 
discontinuity (as may occur inside a pulsar cavity 
inflated by one of the NSs in the precursor binary.)  Below we
discuss other possible causes for such a broad and largely achromatic
peak in the afterglow light curve. The main features these models need
to explain are the extremely low value of $F_X(11\,{\rm hr})/S_\gamma$
and the late-time peak (a few days) in the afterglow light curve.

\noindent {\bf Off-axis jet:} The bulk of the kinetic energy in the
afterglow shock might not be directed along our line of sight, and
could instead point somewhat away from us. For such an off-axis
viewing angle (relative to the region of bright afterglow emission,
envisioned to be a jet of initial half-opening angle $\theta_0$) the
afterglow emission is initially strongly beamed away from us
\citep[this can be thought of as an extreme version of the ``patchy
shell'' model --][]{KP00b,NPG03}. As the afterglow jet decelerates the
beaming cone of its radiation widens, until it eventually reaches our
line of sight, at which point the observed flux peaks and later decays
\citep{Rees1999,Dermer+2000,Granot+2002,Ramirez-Ruiz+2005}.
This interpretation can naturally account for the
dim early afterglow emission (without necessarily implying an
extremely low external density), as well as the rapid decay after the
peak (if our viewing angle from the jet axis is $\theta_{\rm obs}
\gtrsim 2\theta_0$).  The possibility of a slightly off-axis jet is
particularly intriguing given the fact that the initial spike is much
fainter relative to the extended emission in this event (and in GRB\
060614, which also exhibits a late light curve peak) than for most
SGRBs; one may envision a unified short-burst model in which the
short-spike component of the prompt emission is beamed more narrowly
than the component associated with the extended emission.  However,
since a low circumstellar density is no longer needed, there is no
natural means of supressing the early afterglow that should be
created by the extended-emission associated component, and producing
the large ratio of the gamma-ray fluence and early-time X-ray
afterglow flux would require that the gamma-ray emission along our
line of sight is bright and the gamma-ray efficiency is very large
\citep{EG06}.
Regardless of whether the jet is seen off-axis, there is good evidence
that this GRB is significantly collimated, with a decay index $\alpha
> 2$ at late times ($t > 3$~d) in both the optical and X-ray bands.

\noindent {\bf Refreshed shock:} A ``refreshed shock''
\citep{KP00a,R-RMR01,GNP03} is a discrete shell of slow ejecta that
was produced during the prompt activity of the source and catches up
with the afterglow shock at a late time (after it decelerates to a
somewhat smaller Lorentz factor than that of the shell), colliding
with it from behind and thus increasing its energy. This
interpretation also requires a very large gamma-ray efficiency,
($\epsilon_\gamma \gtrsim 95\%$) corresponding to
$\epsilon_\gamma/(1-\epsilon_\gamma) \sim \eta_{\rm k\gamma}^{-1} \gtrsim 30$. 
In this picture, the sharp decay after the peak (at least
as steep as $\sim t^{-2}$) requires that the collision occur after the 
jet-break time.

The rather sparse afterglow data make it hard to distinguish between
these options. Nevertheless, the overall observed behavior can be
reasonably explained as afterglow emission in the context of existing
models for afterglow variability.

\subsection{Constraints on a Mini-Supernova}
\label{sec:minisn}

Under any scenario, the absence of a bright afterglow associated with
GRB\ 080503, together with the late-time optical rise, suggests that a
substantial fraction of this event's energy may be coupled to trans- and
non-relativistic ejecta.  Non-relativistic outflows from the central
engine are sufficiently dense to synthesize heavy isotopes, which may
power transient emission via reheating of the (adiabatically cooled)
ejecta by radioactive decay \citep{LiPaczynski1998}.  Since at most
$\sim 0.1$~M$_{\odot}$ is expected to be ejected from any short GRB
progenitor, the outflow becomes optically thin earlier and traps a
smaller fraction of the decay energy than for a normal SN; these
``mini-SNe'' therefore peak earlier and at fainter magnitudes than
normal SNe.

Current observational limits
\citep{Bloom+2006,Hjorth+2005a,Castro-Tirado+2005,Kann+2008} indicate that any
supernova-like event accompanying an SGRB would have to be over 50 times
fainter (at peak) than normal Type Ia SNe or Type Ic hypernovae, 5
times fainter than the faintest known SNe~Ia or SNe~Ic, and fainter
than the faintest known SNe~II. These limits strongly constrain
progenitor models for SGRBs. Unless SGRBs are eventually found to be
accompanied by telltale emission features like the SNe associated with
LGRBs, the only definitive understanding of the progenitors will come
from possible associations with gravitational wave or neutrino signals.

The most promising isotope to produce bright transient emission is
$^{56}$Ni because its decay timescale of $\sim 6$~d is comparable
to the timescale over which the outflow becomes optically thin.
Compact object mergers, however, are neutron rich and are not expected
to produce large quantities of Ni \citep{Rosswog+2003}. \citet{Metzger+2008b} estimate that
in the best cases only $\leq 10^{-3}$~M$_{\odot}$ of Ni is produced by
outflows from the accretion disk. On the other hand, neutron-rich
material may be dynamically ejected from a NS--NS or a NS--BH merger.
Its subsequent decompression may synthesize radioactive elements
through the $r$ process, whose radioactive decay could power an optical
transient \citep{LiPaczynski1998}.  Material dynamically stripped from
a star is violently ejected by tidal torques through the outer
Lagrange point, removing energy and angular momentum and forming a
large tail. These tails are typically a few thousand kilometers in
size by the end of the disruption event.  Some of the fluid (as much
as a few hundredths of a solar mass) in these flows is often
gravitationally unbound, and could, as originally envisaged by
\cite{LattimerSchramm1976}, undergo $r$-process nucleosynthesis
\citep{rs99,frei99}. The rest will
eventually return to the vicinity of the compact object, with possible
interesting consequences for SGRB late-time emission. 
A significant fraction ($\sim 10$--50\%) of the accretion disk that
initially forms from the merger will also be ejected in powerful winds
\citep{Lee+2005} from the disk at late times; this material is also 
neutron rich and will produce radioactive isotopes \citep{Metzger+2008c}.

In the case of GRB\ 080503, the amount (mass $M$) of radioactive
material synthesized in the accompanying SGRB wind necessary to
provide the observed luminosity is constrained to be $(M/{\rm
M}_\odot)f \approx (1.5 - 1.8) \times 10^{-7}\,(z/1)^2$. A larger
uncertainty is the value of $f$, which is the fraction of the rest
mass of the radioactive material that is converted to heat and
radiated around the optical near the peak of the light curve ($\sim$
1--2~d). Generally $f \lesssim 10^{-4}$ since $\sim 10^{-3}$ of the
rest mass is converted to gamma-rays during the radioactive decay,
only part of the gamma-ray energy is converted to heat (some
gamma-rays escape before depositing most of their energy), and only
part of mass in the synthesized radioactive elements decays near the
peak of the light curve (so that $f$ can easily be much less than
$10^{-4}$, but it is hard for it to be higher than this value).  We
note here that the most efficient conversion of nuclear energy to the
observable luminosity is provided by the elements with a decay
timescale comparable to the timescale it takes the ejected debris to
become optically thin ($t_\tau$). In reality, there is likely to be a
large number of nuclides with a very broad range of decay
timescales. Current observational limits thus place interesting
constraints on the abundances and the lifetimes of the radioactive
nuclides that form in the rapid decompression of nuclear-density
matter --- they should be either very short or very long when compared
to $t_\tau$ so that radioactivity is inefficient in generating a high
luminosity.

\begin{figure}
\centerline{
\includegraphics[scale=0.7,angle=0]{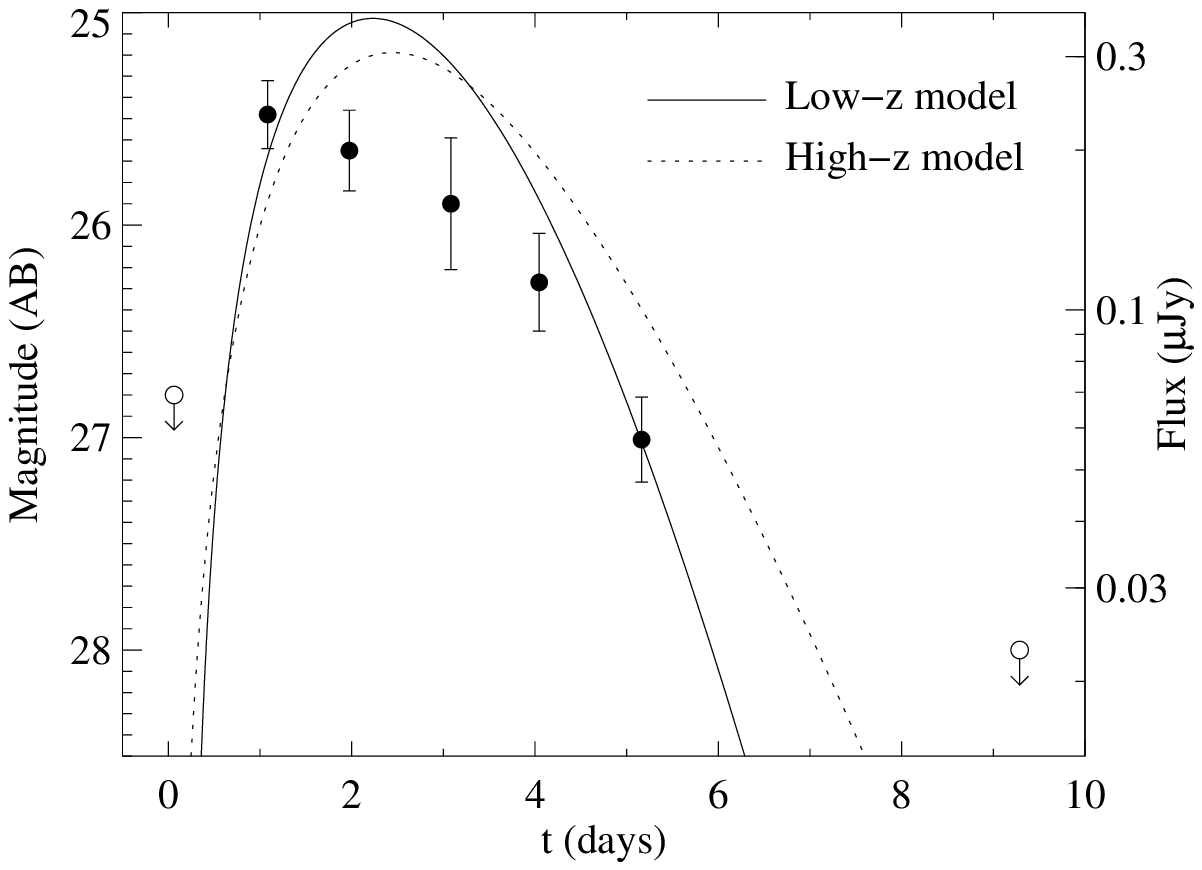}}
\caption{Two AB magnitude \citep{Oke1974} light-curve models for a
Ni-powered ``mini-SN'' from GRB\ 080503, based on the model of
\cite{LiPaczynski1998}, \cite{Kulkarni2005}, and \cite{Metzger+2008b}.
The solid line indicates a model at $z = 0.03$ with a $^{56}$Ni mass
$\approx 2\times 10^{-3}$~M$_{\odot}$, total ejecta mass $\approx
0.4$~M$_{\odot}$, and outflow velocity $\approx 0.1 c$. The dotted
line is for a pure Ni explosion at $z = 0.5$ with mass $\approx
0.3$~M$_{\odot}$ and velocity $\approx 0.2 c$.  Also shown are our
$r$-band and F606W detections and upper limits from Gemini and {\it
HST}.}
\label{fig:minisn}
\end{figure}

In Figure \ref{fig:minisn} we show two light-curve models for a
Ni-powered mini-SN from GRB\ 080503 calculated according to the
model of \cite{Kulkarni2005} and \cite{Metzger+2008b}.  Shown with
asterisks and triangles are the $r$-band and F606W band detections and
upper limits from Gemini and {\it HST}.  The solid and dashed lines
correspond to a low-redshift ($z = 0.03$) and high-redshift ($z =
0.5$) model, respectively.  Qualitatively, both models appear to be
reasonably consistent with the flux light curve. To reproduce the peak
of the optical emission at $t \approx 1$~d, a total ejected mass of
$\sim 0.1$~M$_{\odot}$ is required in either model.  In order to
reproduce the peak flux, the Ni mass required in the high- and 
low-redshift models is $M_{\rm Ni} \approx 0.3$~M$_\odot$ and $2 \times
10^{-3}$~M$_{\odot}$, respectively.  Since the former is unphysically large
in any SGRB progenitor model, a high-redshift event appears
inconsistent with a mini-SN origin for the optical rise.

If GRB\ 080503 originates at very low redshift ($z < 0.1$), a mini-SN
model would still appear viable.  However, most mini-SN models also
predict that the spectrum should redden significantly with time and
possess a negative spectral slope once the outflow becomes optically
thin after the peak at $t \approx 1$~d; the {\it HST} detection in
F606W and non-detections in F814W and F450W at 5.35~d, however,
suggest that the spectrum is approximately flat at late times.  While
the detected optical emission may be attributed to a mini-SN type of
event, the expected spectrum in such a case is quasi-thermal,
resulting in no detectable emission in the X-rays.
(\citealt{Rossi+2008} have proposed a fallback model in which X-rays can
rebrighten days or weeks after the event, but the luminosity is
extremely low, and to explain the Chandra count rate a very close
distance of $\sim 8$~Mpc would be required; while not
excluded by our data, this is orders of magnitude closer than any known
non-magnetar short gamma-ray burst.)  Therefore, the late X-ray
detections a few days after the GRB are most likely afterglow
emission.

\section{Conclusions}
\label{sec:conclusions}

The very same faintness which makes GRB\ 080503 so remarkable 
unfortunately also makes it difficult to strongly constrain
various physical interpretations of this event.  However, the
combination of the extremely low limit on the afterglow-to-prompt
fluence ratio shortly after the burst and the lack of a coincident
host galaxy provides strong evidence that this burst exploded in a
very low-density (possibly even intergalactic) medium.

This result has several important implications for the nature of
``short'' bursts and of GRB classification in general.  For example,
the interpretation of GRB\ 060614 (and whether it groups more
naturally with canonical ``short'' events like GRB\ 050724, canonical
``long'' events like 080319B, or in a new class entirely on its own)
is clarified somewhat.  GRB\ 060614, despite having a prompt-extended
light-curve morphology (as well as negligible lag and no supernova to
deep limits) was (like GRB\ 080503) strongly dominated by extended
emission but also had a very long spike $T_{90}$ (5.5~s), on the
extreme end of the short class.  The initial pulse of GRB\ 080503 was
unambiguously short; furthermore, the faint afterglow and lack of host
galaxy both provide evidence that this event occurred in an
environment quite unlike those of canonical ``long'' GRBs.  The
existence of an apparent continuity between the appearance of the
light curves of GRB\ 060614 and GRB\ 080503 and more traditional short
bursts (in stark contrast to the bewildering diversity in the
structure of longer GRBs) suggests that they originate from the same
or similar progenitors, in spite of the apparent diversity in
environments and redshifts.  The presence of bright extended emission
in GRB\ 080503, and the prompt-like behavior of its fading tail in the
X-ray band, is a counterexample to the inference that extended
emission is an environment- or progenitor-correlated phenomenon
\citep{Troja+2008}.  We note again that in the vast majority of cases
observed by {\it Swift}, we cannot strongly constrain the presence of
extended emission, and in only two events are limits sufficiently deep
to constrain the extended-to-spike fluence ratio to less than the value
observed for GRB\ 070714B.

This same result, however, may pose difficulties to the most popular
model of short GRBs: NS--NS or NS--BH merger events.  The possibility
that the luminosity of the extended emission can exceed that of the
initial spike by factors of 30 or more is problematic for a merger, in
which the majority of the accretion disk is expected to accrete within
a viscous timescale --- not more than a few seconds
\citep{Rosswog+2007,Lee+2004}.  This may strengthen the case for alternative
models, such as accretion-induced collapse \citep{VietriStella1999,KatzCanel1996,MacFadyen+2005}.
On the other hand, the
extremely low circumburst density is much more consistent with a
merger event with its possibility of a natal kick than models such as
accretion-induced collapse. One possible means of avoiding this
difficulty in a merger scenario (but which could also apply to other
models) would be if, for GRB\ 080503 and GRB\ 060614, the prompt spike
were focused in a narrow jet seen nearly off-axis while the extended
emission were more widely beamed.  Such a scenario could occur in the 
case of compact object mergers if the relativistic jet is collimated 
by a neutrino-heated baryon wind from the accretion disk at early times 
\citep{LevinsonEichler2000,Rosswog+2003}, but the collimating effect of the wind
become less effective at later times as the neutrino flux and wind
luminosity decreases.

The observed late peak in the optical light curve, which we suspected
initially may have been the signature of a Li-Paczy\'{n}ski supernova, is
explained reasonably by other models.  The peak time of $\sim$1~d is
too long to be explained by the deceleration timescale, even for a
burst exploding into the extremely low-density intergalactic medium,
unless the Lorentz factor associated with the extended episode is
also very low.  However, an off-axis jet, or alternatively 
a slower shell of ejecta that catches up with the initially very weak
afterglow shock and energizes it (a ``refreshed shock''), could
produce a rebrightening and a late peak. A rather similar late peak
has been observed before in several long bursts and in GRB 060614.  
Some contribution to the afterglow from
a mini-SN is not ruled out but is not necessary to explain the
available data.

Our failure to conclusively detect a mini-SN signature may also have
significant observational implications.  In spite of the ``nakedness''
of this event vastly suppressing the late-time afterglow flux, any
possible mini-SN that may have been associated with this event was
concealed by the late-time afterglow.  Similar events in a
higher-density environment (such as a galactic disk) will have even
brighter afterglows.  If mini-SN phenomena exist in nature, our
observations suggest it will be extremely difficult to detect them
over the glow of the relativistic shock created by the burst itself.
Our best opportunity is likely to lie in observationally and
intrinsically faint events like GRB\ 050509B, whose weak gamma-ray signal
results from a low-energy flow insufficient to create a bright afterglow
even in a relatively dense medium, but is bright enough for localization.  

\acknowledgements

J.S.B.'s group is supported in part by the Hellman Faculty Fund, Las
Cumbres Observatory Global Telescope Network, and NASA/\textit{Swift}
Guest Investigator grant NNG05GF55G.  BM and EQ were supported in 
part by the David and Lucile Packard Foundation, NASA grant NNG05GO22H,
and the NSF-DOE Grant PHY-0812811.  J.G. gratefully acknowledges a Royal 
Society Wolfson Research Merit Award.  N.R.B. is partially supported by 
US Department of Energy SciDAC grant DE-FC02-06ER41453 and by a NASA 
GLAST/Fermi Fellowship.  A.V.F. is
partially supported by NSF grant AST--0607485 and the TABASGO Foundation.
This work was supported in part by NASA (Swift NX07AE98G, ER-R) and DOE 
SciDAC (DE-FC02-01ER41176, ER-R).

This research is based in part on observations obtained at the Gemini
Observatory, which is operated by the Association of Universities for
Research in Astronomy, Inc., under a cooperative agreement with the
NSF on behalf of the Gemini partnership.
Some of the data presented herein were obtained at the W. M. Keck
Observatory, which is operated as a scientific partnership among the
California Institute of Technology, the University of California, and
the National Aeronautics and Space Administration (NASA). The
Observatory was made possible by the generous financial support of the
W. M. Keck Foundation.

We thank the {\it HST} and Chandra X-ray Observatory directors and
scheduling teams for their extremely rapid turnaround time for
observations of GRB\ 080503.  We also thank the Gemini observing staff,
in particular T. Geballe, for excellent support, and D. A. Kann 
for helpful commentary on the manuscript.

\bibliographystyle{apj}
\bibliography{ref}

\clearpage

\begin{deluxetable}{llll}                                               
\tablecaption{Prompt Emission Properties of Swift SGRBs and Candidate SGRBs} 
\tablehead{ \colhead{GRB} & \colhead{ambiguous?} & \colhead{$z$} &
\colhead{$S_{\rm EE}/S_{\rm spike}$} \\
}                                               
\startdata                                               
050509B & N                  & 0.2249 &$ < 14.3 $ \\ 
050724  & N                  & 0.258  &$ 2.64 \pm 0.49 $ \\
050813  & N                  & 0.722? &$ < 3.64 $ \\
050906  & Y\tablenotemark{a} & -      &$ < 14.87 $ \\ 
050911  & Y\tablenotemark{b}\tablenotemark{c}& 0.1646? &$ 1.31 \pm 0.43 $ \\
050925  & Y\tablenotemark{d} & -      &$ < 1.83 $ \\
051105A & N                  & -      &$ < 8.06 $\\
051210  & Y\tablenotemark{b} & 0.114? &$ 2.72 \pm 1.33$ \\ 
051221A & Y\tablenotemark{b} & 0.5465 &$ < 0.16 $ \\ 
051227  & Y\tablenotemark{b} & -      &$ 2.87 \pm 0.677 $ \\ 
060313  & N                  & -      &$ < 0.29 $ \\ 
060502B & N                  & 0.287? &$ < 3.45 $ \\ 
060801  & N                  & 1.131? &$ < 1.84 $ \\ 
060614  & Y\tablenotemark{b}\tablenotemark{e} & 0.125 &$ 6.11 \pm 0.25 $ \\ 
061006  & Y\tablenotemark{b} & 0.4377  &$ 1.75 \pm 0.26 $ \\ 
061201  & N                  & 0.111? &$ < 0.71 $ \\ 
061210  & N                  & 0.41?  &$ 2.81 \pm 0.63 $ \\ 
061217  & N                  & 0.827  &$ < 3.81 $ \\ 
070209  & N                  & -      &$ < 8.08 $ \\ 
070429B & N                  & 0.904 &$ < 2.44 $ \\ 
070714B & N                  & 0.92   &$ 0.477 \pm 0.163 $ \\ 
070724A & N                  & 0.457  &$ < 4.24 $ \\ 
070729  & N                  & -      &$ < 2.16 $ \\ 
070731  & Y\tablenotemark{b} & -      &$ < 1.37 $ \\ 
070809  & Y\tablenotemark{b} & 0.219? &$ < 1.37 $ \\ 
070810B & N                  & -      &$ < 9.40 $ \\ 
070923  & N                  & -      &$ < 5.96 $ \\ 
071112B & N                  & -      &$ < 4.14 $ \\ 
071227  & Y\tablenotemark{b} & 0.383  &$ 1.56 \pm 0.49 \tablenotemark{f} $ \\ 
080503  & Y\tablenotemark{e} & -      &$ 32.41 \pm 5.7 $ \\ 
\enddata                                                
\tablenotetext{a}{SGR flare in IC 328?}                                                
\tablenotetext{b}{Spike $T_{90} > 1$~s.}                                                
\tablenotetext{c}{Extended-emission episode is of much shorter duration 
than in all other events.}
\tablenotetext{d}{Soft event; in Galactic plane.}
\tablenotetext{e}{Fluence dominated by extended emission.}
\tablenotetext{f}{Significance of the extended emission is $<4\sigma$.}
\label{tab:battable}                                                
\end{deluxetable}                                                                                                  

\begin{deluxetable}{llcllll}
\tablecaption{Optical and Near-IR Observations of the Optical 
Counterpart of GRB 080503}
\tablehead{
\colhead{$t_{\rm mid}$} & \colhead{ Exp. time } & \colhead{ filter } &
\colhead{ magnitude } & \colhead{$\lambda$} & \colhead{flux (or
limit)} & \colhead{telescope} \\
\colhead{(day)} & \colhead{(s)} & \colhead{} & \colhead{} & \colhead{(\AA)}
& \colhead{($\mu$Jy)} & \colhead{} }
\startdata 
0.00156 & 98 & white& $> 20            $ & 3850  & $ < 14.2             $ & Swift UVOT   \\
0.04083 &180 & r    & $> 25.80         $ & 6290  & $ < 0.204            $ & Gemini-N GMOS\\
0.04916 &800 & g    & $  26.76 \pm 0.24$ & 4858  & $  0.0890 \pm 0.0176 $ & Gemini-N GMOS\\ 
0.06250 &800 & r    & $> 26.80         $ & 6290  & $ < 0.0811           $ & Gemini-N GMOS\\
0.05125 &300 & B    & $> 26.00         $ & 4458  & $ < 0.209            $ & Keck I LRIS\\
0.05458 &630 & R    & $> 25.60         $ & 6588  & $ < 0.208            $ & Keck I LRIS\\
0.07583 &800 & i    & $> 26.80         $ & 7706  & $ < 0.0779           $ & Gemini-N GMOS\\
0.09000 &800 & z    & $> 26.00         $ & 9222  & $ < 0.161            $ & Gemini-N GMOS\\
0.10125 &360 & g    & $> 24.60         $ & 4858  & $ < 0.650            $ & Gemini-N GMOS\\
1.08333 &1800& r    & $  25.48 \pm 0.16$ & 6290  & $  0.273 \pm 0.037   $ & Gemini-N GMOS\\
1.97500 &1620& r    & $  25.65 \pm 0.19$ & 6290  & $  0.234 \pm 0.038   $ & Gemini-N GMOS\\
2.09167 & 720& g    & $  26.48 \pm 0.26$ & 4858  & $  0.115 \pm 0.024   $ & Gemini-N GMOS\\
3.08333 &2700& r    & $  25.90 \pm 0.31$ & 6290  & $  0.186 \pm 0.046   $ & Gemini-N GMOS\\
4.04583 &2880& r    & $  26.27 \pm 0.23$ & 6290  & $  0.132 \pm 0.025   $ & Gemini-N GMOS\\
5.20833 &2760& $K_s$& $> 22.47         $ & 21590 & $ < 0.700            $ & Gemini-N NIRI\\
5.35833 &4600& F606W& $  27.01 \pm 0.20$ & 6000  & $  0.067 \pm 0.011   $ & HST WFPC2\\
5.35833 &2100& F450W& $> 26.9          $ & 4500  & $ < 0.080            $ & HST WFPC2\\
5.35833 &2100& F814W& $> 26.8          $ & 8140  & $ < 0.077            $ & HST WFPC2\\
9.12917 &4000& F814W& $> 27.1          $ & 6000  & $ < 0.058            $ & HST WFPC2\\
9.12917 &4000& F606W& $> 28.0          $ & 6000  & $ < 0.027            $ & HST WFPC2\\
\enddata
\tablecomments{SDSS magnitudes are given in AB, while $B$ and $R$ are under the Vega system.  $K_s$ is relative to the 2MASS system \citep{2MASScalib}.  Flux values given are corrected for foreground
extinction ($E_{B-V} = 0.06$, \citealt{Schlegel+1998}) while magnitudes are uncorrected.  Limits are $3 \sigma$ values.}
\label{tab:photometry}
\end{deluxetable}

\begin{deluxetable}{llllllllll}
\tablecaption{Magnitudes of Faint Secondary Standards in the GRB~080503 Field}
\tablehead{
\colhead{RA} & \colhead{dec} & \colhead{$g$} & \colhead{$r$} &
\colhead{$i$} & \colhead{$z$} & \colhead{$B$} & \colhead{$V$} &
\colhead{$R$} & \colhead{$I$}\\ \colhead{(hh:mm:ss)} &
\colhead{(dd:mm:ss)} & \colhead{(mag)} & \colhead{(mag)} & \colhead{(mag)} &
\colhead{(mag)} & \colhead{(mag)} & \colhead{(mag)} & \colhead{(mag)} &
\colhead{(mag)} }
\startdata 
19:06:16.785 & +68:46:41.39 & 19.890 & 18.677 & 18.116 & 17.611 & 20.496 & 19.185 & 18.363 & 17.562 \\
19:06:27.931 & +68:46:55.62 & 21.338 & 20.790 & 20.671 & 20.410 & 21.736 & 21.017 & 20.602 & 20.217 \\
19:06:40.791 & +68:47:14.20 & 20.894 & 20.412 & 20.324 & 20.103 & 21.272 & 20.612 & 20.235 & 19.881 \\
19:06:47.096 & +68:47:44.06 & 20.394 & 19.308 & 19.008 & 18.622 & 20.961 & 19.762 & 19.044 & 18.508 \\
19:06:25.306 & +68:48:47.91 & 18.552 & 17.882 & 17.776 & 17.457 & 18.988 & 18.161 & 17.685 & 17.313 \\
19:06:31.664 & +68:48:32.11 & 19.158 & 17.851 & 17.349 & 16.847 & 19.794 & 18.398 & 17.537 & 16.803 \\
19:06:25.808 & +68:47:18.09 & 21.583 & 20.428 & 20.040 & 19.625 & 22.171 & 20.911 & 20.145 & 19.524 \\
19:06:33.303 & +68:48:01.76 & 22.115 & 20.725 & 19.555 & 18.840 & 22.777 & 21.307 & 20.305 & 18.887 \\
19:06:42.332 & +68:48:05.17 & 18.885 & 18.453 & 18.412 & 18.187 & 19.247 & 18.632 & 18.287 & 17.974 \\
19:06:26.337 & +68:46:57.77 & 23.241 & 21.868 & 21.077 & 20.510 & 23.898 & 22.443 & 21.506 & 20.483 \\
19:06:42.896 & +68:48:08.70 & 21.778 & 21.238 & 21.066 & 20.696 & 22.174 & 21.462 & 21.043 & 20.585 \\
19:06:38.937 & +68:47:44.69 & 19.863 & 19.421 & 19.361 & 19.129 & 20.228 & 19.604 & 19.251 & 18.920 \\
19:06:29.508 & +68:47:49.97 & 23.138 & 21.832 & 20.560 & 19.793 & 23.774 & 22.379 & 21.405 & 19.870 \\
19:06:39.266 & +68:47:48.01 & 19.900 & 19.085 & 18.872 & 18.542 & 20.382 & 19.425 & 18.859 & 18.394 \\
19:06:34.192 & +68:46:35.61 & 19.806 & 19.179 & 19.040 & 18.756 & 20.229 & 19.440 & 18.981 & 18.579 \\
19:06:33.173 & +68:46:33.32 & 20.870 & 20.333 & 20.208 & 19.946 & 21.265 & 20.556 & 20.145 & 19.753 \\
19:06:29.230 & +68:46:10.00 & 22.166 & 20.792 & 19.444 & 18.661 & 22.823 & 21.368 & 20.348 & 18.741 \\
19:06:29.556 & +68:46:12.79 & 23.156 & 22.015 & 21.577 & 21.144 & 23.740 & 22.492 & 21.726 & 21.052 \\
19:06:18.146 & +68:47:56.94 & 23.169 & 21.896 & 21.373 & 20.906 & 23.794 & 22.429 & 21.582 & 20.831 \\
19:06:29.343 & +68:46:23.23 & 24.127 & 22.729 & 21.678 & 21.039 & 24.791 & 23.315 & 22.326 & 21.039 \\
19:06:37.946 & +68:48:28.37 & 24.313 & 23.088 & 21.740 & 20.917 & 24.923 & 23.601 & 22.657 & 21.030 \\
19:06:30.838 & +68:48:06.34 & 24.127 & 22.581 & 21.503 & 20.803 & 24.838 & 23.229 & 22.161 & 20.849 \\
19:06:21.135 & +68:48:18.82 & 24.853 & 23.538 & 22.316 & 21.578 & 25.491 & 24.089 & 23.118 & 21.637 \\ 
\enddata
\tablecomments{Magnitudes of calibration stars in the field of GRB\
080503 as measured using repeated observations of the SA~110 field at
varying airmasses over four photometric nights.  Uncertainties in all
cases are dominated by color terms and are approximately 0.02--0.05
mag.}
\label{tab:calibstars}
\end{deluxetable} 

\clearpage

\end{document}